\documentclass{article}

\usepackage{PRIMEarxiv}

\usepackage[utf8]{inputenc} % allow utf-8 input
\usepackage[T1]{fontenc}    % use 8-bit T1 fonts
\usepackage{url}            % simple URL typesetting
\usepackage{booktabs}       % professional-quality tables
\usepackage{amsfonts}       % blackboard math symbols
\usepackage{nicefrac}       % compact symbols for 1/2, etc.
\usepackage{microtype}      % microtypography
\usepackage{lipsum}
\usepackage{fancyhdr}       % header
\usepackage{graphicx}       % graphics
\graphicspath{{media/}}     % organize your images and other figures under media/ folder
\usepackage{float}
\usepackage{tikz}
\usetikzlibrary{mindmap}
\usepackage{graphicx}
\usepackage{amsmath}
\usepackage{array}
\usepackage{amssymb}
\usepackage{tabularx}
\usepackage{booktabs}
\usepackage{xcolor}
\usepackage{color}
\usepackage[english]{babel}
\usepackage{amsthm}
\usepackage{multirow}
\usetikzlibrary{arrows.meta}
\usepackage{tikz}
\usetikzlibrary{shapes,arrows,positioning}
\usepackage{stfloats}
\usepackage{makecell}
\usepackage{adjustbox}
\usepackage{float}
\usetikzlibrary{arrows.meta,positioning,shapes.geometric}
\usepackage{etoolbox}
\usepackage[utf8]{inputenc}
\usepackage{blindtext}
\usepackage{listings}
\usepackage{pifont}
\usepackage{wasysym}
\usepackage{lipsum}
\usepackage{appendix}

\usepackage{array}
\newcolumntype{C}[1]{>{\centering\arraybackslash}p{#1}}
\theoremstyle{definition}

\newcolumntype{L}[1]{>{\raggedright\arraybackslash}p{#1}}

\newtheorem{theorem}{Theorem}

\definecolor{ta-safe}{HTML}{6BAED6}        % Safe Autonomy
\definecolor{ta-perm}{HTML}{74C476}        % Permissible Zone (Low-trust -> Low-risk)
\definecolor{ta-hitl}{HTML}{FDBE6F}        % HITL Zone
\definecolor{ta-fail}{HTML}{E34A33}        % Failure State
\definecolor{ta-attack}{HTML}{F16913}

\lstdefinelanguage{Policy}{
  morekeywords={policy,rule,when,then,allow,deny,require,exists,forall,if,and,or,not,in,return},
  sensitive=true, morecomment=[l]{//}, morestring=[b]"
}
\newcolumntype{M}[1]{>{\centering\arraybackslash}m{#1}}

\title{SoK: Trust-Authorization Mismatch in LLM Agent Interactions}

\author{
  Guanquan Shi\\
  \texttt{shiguanquan@buaa.edu.cn} \\
   \And
  Haohua Du* \\
  \texttt{duhaohua@buaa.edu.cn} \\
   \And
   Zhiqiang Wang \\
  \texttt{sa21221041@mail.ustc.edu.cn} \\
   \And
   Xiaoyu Liang \\
  \texttt{xiaoyuliang@buaa.edu.cn} \\
   \And
   Weiwenpei Liu \\
  \texttt{liuweiwenpei@buaa.edu.cn} \\
   \And
   Song Bian \\
  \texttt{sbian@buaa.edu.cn} \\
   \And
   Zhenyu Guan\\
  \texttt{guanzhenyu@buaa.edu.cn} \\
}

\begin{document}
\maketitle

\begin{abstract}
Abstract Large Language Models (LLMs) are evolving into autonomous agents capable of executing complex workflows via standardized protocols (e.g., MCP). However, this paradigm shifts control from deterministic code to probabilistic inference, creating a fundamental Trust-Authorization Mismatch: static permissions are structurally decoupled from the agent's fluctuating runtime trustworthiness. In this Systematization of Knowledge (SoK), we survey more than 200 representative papers to categorize the emerging landscape of agent security. We propose the Belief-Intention-Permission (B-I-P) framework as a unifying formal lens. By decomposing agent execution into three distinct stages—Belief Formation, Intent Generation, and Permission Grant—we demonstrate that diverse threats, from prompt injection to tool poisoning, share a common root cause: the desynchronization between dynamic trust states and static authorization boundaries. Using the B-I-P lens, we systematically map existing attacks and defenses and identify critical gaps where current mechanisms fail to bridge this mismatch. Finally, we outline a research agenda for shifting from static Role-Based Access Control (RBAC) to dynamic, risk-adaptive authorization.

\end{abstract}

% keywords can be removed
\keywords{AI agent \and trustworthiness \and security}

\section{Introduction}
\label{sec1:intro}

\begin{figure}[t]
    \centering
    \includegraphics[width=.6\textwidth]{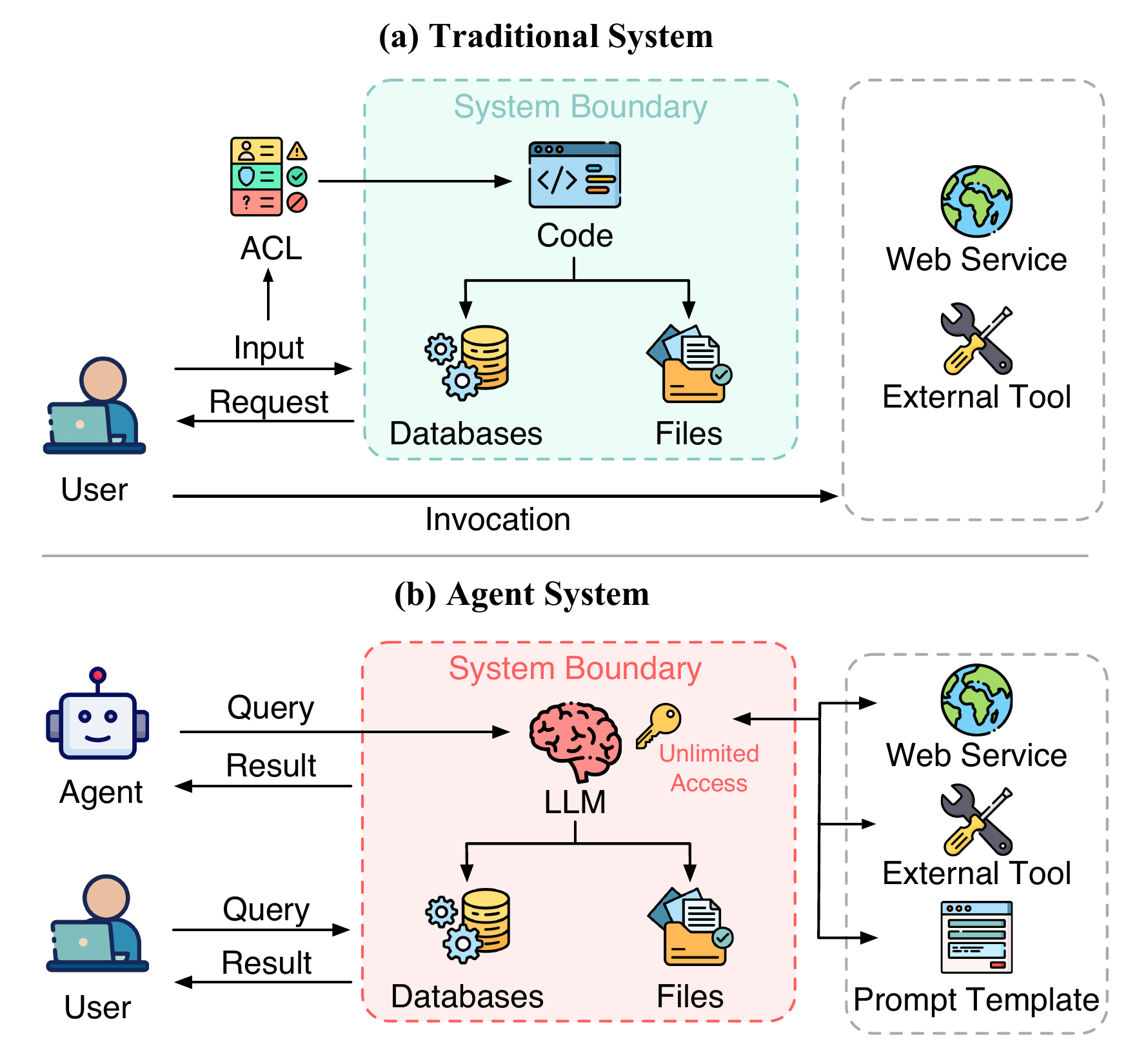}
    \caption{The Comparison of the Traditional Security Model and AI Trustworthy Model.}
    \label{fig:intro-comp}
\end{figure}

LLMs are transcending their roles as passive information retrieval engines to become autonomous agents capable of executing complex workflows in digital and physical environments~\cite{xie2024large}. Through standardized protocols such as the Model Context Protocol (MCP)~\cite{anthropic2024introducing}, these agents now have the ability to plan tasks, invoke external tools, and manipulate system resources. As illustrated in Fig.~\ref{fig:intro-comp}, this evolution fundamentally shifts the locus of control from deterministic code paths—manually crafted and statically verifiable—to probabilistic, natural language-driven decision models. While this paradigm shift unlocks immense utility, it destabilizes the foundational assumptions of modern system security: that software behavior is predictable and that privileges granted at authentication remain valid throughout execution~\cite{latham1986department}.

Specifically, incumbent security architectures rely on pre-defined policy evaluation~\cite{john2025owasp}. Systems typically perform a one-time entitlement check upon authentication, binding a fixed set of privileges to an agent's identity. This model assumes that an authenticated entity will strictly adhere to defined business logic. However, LLM agents defy this assumption. Governed by stochastic sampling rather than deterministic logic, an agent's runtime behavior is highly context-dependent. According to Rice’s Theorem~\cite{rice1953classes}, we analyze that verifying whether a Turing-complete agent's semantic behavior satisfies a non-trivial security property for arbitrary inputs is undecidable before runtime. 
\textit{Consequently, static access control mechanisms are structurally incapable of governing probabilistic agent behaviors.}
We term this structural misalignment the \textbf{Trustworthiness-Authorization Mismatch}. It represents a systemic failure state in which the static execution permissions granted to an agent become dangerously decoupled from the agent's fluctuating trustworthiness at runtime. 
In this state, a compromised or hallucinating agent retains high-privilege access (Authorization) even when its internal belief state or immediate intent (Trustworthiness) has been corrupted.

Despite the escalating threat landscape, the academic community’s understanding of this domain remains fragmented. Existing taxonomies often focus on listing specific attack vectors, such as ``Jailbreaking" or ``Prompt Injection," without explaining the underlying structural failures that allow these attacks to generalize across different modalities and protocols. Without a unified framework to analyze the agent's decision lifecycle, defenses remain reactive and isolated.

% This Systematization of Knowledge (SoK) bridges the widening gap between system security and AI alignment~\cite{greenblatt2024alignment,bommasani2023holistic}. 
% We argue that, without resolving the underlying Trustworthiness-Authorization Mismatch, relying solely on AI alignment is not only brittle (susceptible to adversarial attacks) but also detrimental to utility due to excessive refusals (effectively a denial-of-service attack). Similarly, relying solely on boundary isolation fails to prevent legitimate agents from being induced to perform malicious operations (the Confused Deputy problem).

% Thus, we formalize the B-I-P (Belief-Intention-Permission) framework. This model serves as a rigorous lens to critically re-examine the attack and defense literature of the past two years. By decomposing agentic security into these three primitives, we demonstrate that disparate incidents—from prompt injection to tool poisoning—share a common root cause: the system's failure to dynamically modulate \emph{Permission} based on the agent's real-time \emph{Belief} state, leading to unauthorized \emph{Intentions}.

This Systematization of Knowledge (SoK) provides a unifying formal lens for agent interaction security. We argue that addressing the security of autonomous agents requires shifting from a ``vulnerability-patch" mindset to a ``lifecycle-state" perspective. To this end, we formalize the Belief-Intention-Permission (B-I-P) framework. This framework decomposes the opaque "black box" of agent execution into three observable stages: (1) Belief Formation, where the agent constructs a probabilistic view of the world from inputs; (2) Intent Generation, where beliefs are synthesized into provisional plans; and (3) Permission Grant, where plans encounter the hard boundary of system authorization.

By grounding our analysis in the B-I-P framework, we shift from a phenomenological account of attacks to a mechanistic understanding of failure. We surveyed 87 representative papers (screened from 248 candidates) across AI and systems security venues. We demonstrate that disparate incidents—from indirect prompt injection in RAG systems to tool poisoning in MCP servers—share a common root cause: the desynchronization between the agent's dynamic trust state and the system's static authorization boundaries.

\textbf{Scope and Threat Model.} We focus strictly on \emph{runtime interaction security} for agentic systems. We exclude training-time poisoning unless it directly manifests as a runtime decision vulnerability. Our threat model assumes a highly capable adversary who can manipulate user prompts, external retrieval content (RAG), and protocol metadata, targeting the integrity and availability of the $B \!\to\! I \!\to\! P$ decision chain.

% \textbf{Why SoK?} While existing taxonomies catalog \emph{what} attacks look like (e.g., classifying types of prompt injections), they often fail to explain \emph{why} these failures generalize across different modalities and protocols. By grounding the field in the B--I--P model, we transition from a phenomenology of attacks to a mechanistic understanding of failure. This enables us to (1) provide a unified vocabulary connecting isolated findings in AI and Security venues, (2) identify structural ``breakpoints" where the trust chain is severed, and (3) propose testable invariants for future defense designs.

To summarize, our contributions are:
\begin{itemize}
    \item \textbf{A Unifying Formal Lens (The B-I-P Framework)}: We introduce the Belief-Intention-Permission framework to model the agentic lifecycle. We propose the \textit{Safety Margin Ratio (SMR)} as a governing invariant, defining security not as a static property but as a dynamic state in which the system's Trust Capacity must strictly dominate the Risk Load of the intended action ($T(B,I) > R(P)$).

    \item \textbf{Systematization of Attacks \& Defenses}: We provide a comprehensive mapping of existing literature onto the B-I-P lifecycle. Unlike prior surveys that categorize work by attack modality (e.g., ``Image Attacks"), we classify work by its chain-breaking mechanism—distinguishing, for instance, between defenses that verify Data Provenance in Stage 1 and those that enforce Context-Aware Isolation in Stage 3.
    
    % \item \textbf{Systematization of Attacks \& Defenses.} We assemble and code a comprehensive corpus across system security and AI venues. We re-map existing defenses not by their implementation (e.g., "filtering") but by their \emph{chain-breaking mechanism} (e.g., preventing $B \to I$ corruption vs. blocking $I \to P$ execution), revealing critical coverage gaps.
    % \item \textbf{Design Principles for Agentic Security.} We synthesize our findings into a set of architectural requirements for next-generation agents, advocating for \emph{belief-aware authorization} and \emph{intent provenance}. We outline a concrete research agenda to move beyond static RBAC toward dynamic, risk-adaptive access control protocols.
    \item \textbf{Gap Analysis \& Research Agenda}: Our systematization reveals that current defenses are skewed towards the early stages (Belief/Intent) while neglecting the critical bottleneck of Authorization. We outline a concrete research agenda to move beyond static Role-Based Access Control (RBAC) toward Belief-Aware Access Control (BAAC), advocating for authorization mechanisms that can dynamically modulate permissions based on the agent's real-time cognitive uncertainty and data provenance.
    
\end{itemize}

% \textcolor{red}{\textbf{Paper Organization.}
% Sec.~2 provides background, terminology, and our methodology/scope. Sec.~3 introduces the B--I--P model and the Trust--Authorization Matrix/Process. Sec.~4 systematizes attacks via this lens; Sec.~5 systematizes defenses as \emph{chain-breaking} mechanisms; Sec.~6 outlines opportunities and open challenges; Sec.~7 concludes. }
\section{Problem Overview \& Scope}
\label{sec:overview}

% In this section, we analyze the fundamental incompatibility between the operational nature of LLM-based Agents and contemporary access control paradigms. 
% We first establish the non-deterministic nature of agent workflows, then provide formal proofs demonstrating the failure of static and context-aware permission models.
This section characterizes the fundamental conflict between stochastic agent workflows and traditional permission grants, proving that neither static nor context-aware models offer sufficient security guarantees in non-deterministic environments.

% \begin{figure}[t]
%     \centering
%     \includegraphics[width=.5\textwidth]{figue/workflow.pdf}
%     \caption{Execution workflow of the agent system integrated with MCP and A2A protocols.}
%     \label{fig:workflow}
% \end{figure}

% \subsection{Overview of Agent}
We adopt a minimal agent vocabulary consistent with our later formalization: an \emph{agent} maintains internal state, perceives its environment through inputs and tool/agent returns, forms \emph{intentions} to act, and issues actions that are either \emph{permitted} or denied by the host system~\cite{jabbour2024generative}. 
In intent-driven workflows, users express goals in natural language; the agent decomposes goals into subtasks, invokes external capabilities (tools/APIs/agents)~\cite{patil2024gorilla}, and integrates results into a coherent response. 

\subsection{Overview of Agent}
We adopt a minimal agent vocabulary consistent with our later formalization: an \emph{agent} maintains internal state, perceives its environment through inputs and tool/agent returns, forms \emph{intentions} to act, and issues actions that are either \emph{permitted} or denied by the host system~\cite{jabbour2024generative}. 
In intent-driven workflows, users express goals in natural language; the agent decomposes goals into subtasks, invokes external capabilities (tools/APIs/agents)~\cite{patil2024gorilla}, and integrates results into a coherent response. 
This operational loop---from \emph{belief} (about the world) to \emph{intention} (plans) to \emph{permission} (what the system allows) to \emph{action}---is the substrate for our mismatch analysis in Sec.~3.

\begin{figure}[th]
    \centering
    \includegraphics[width=.6\textwidth]{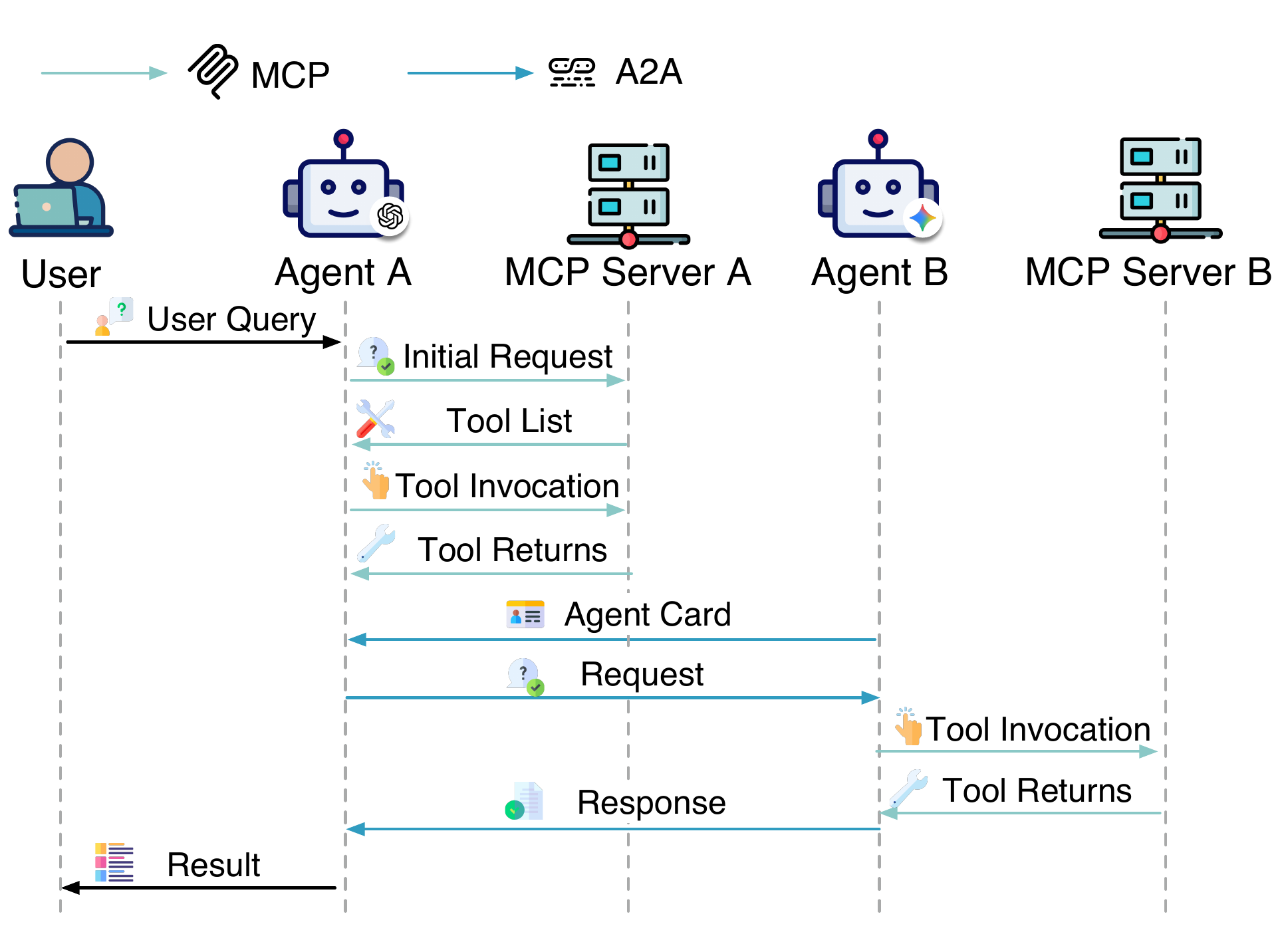}
    \caption{MCP protocol and A2A protocol workflow.}
    \label{fig:workflow}
\end{figure}

\textbf{Interaction Protocols (MCP/A2A) and Boundaries}.
Standardized protocols provide the communication fabric for agent ecosystems. 
An interaction typically proceeds in three stages (cf. Fig.\ref{fig:workflow}):
\begin{enumerate}
  \item \textbf{Capability discovery.} Agents advertise or retrieve metadata describing callable functions and constraints (e.g., MCP tool descriptors; A2A agent cards). This establishes \emph{provenance} and delineates \emph{authorization boundaries}.
  \item \textbf{Invocation and execution.} The initiator supplies parameters and context; calls may nest or cascade to downstream tools/agents. Transports (JSON-RPC, HTTP, etc.) preserve call--return semantics.
  \item \textbf{Result propagation.} Targets return structured outputs (Results/Messages/Artifacts) synchronously or asynchronously; these outputs update the agent's belief state and may trigger further planning.
\end{enumerate}

\textbf{Trust/permission boundaries and instrumentation points.}
To make protocols actionable for security engineering, we annotate the sequence with: 
(i) \emph{source labeling} and \emph{integrity tags} at ingress (where sanitizers and anomaly detectors operate), 
(ii) \emph{belief attribution} at the planner (for quarantine/taint propagation), 
(iii) \emph{policy guards} at permission checks (static/dynamic authorization, belief-aware ABAC/TBAC), and 
(iv) \emph{secure logging} at each transition (for auditability and post-hoc reconstruction). 
These are the concrete \emph{chain-breaking} control points revisited in Sec.~5.

\begin{figure}[t]
    \centering
    \includegraphics[width=.6\textwidth]{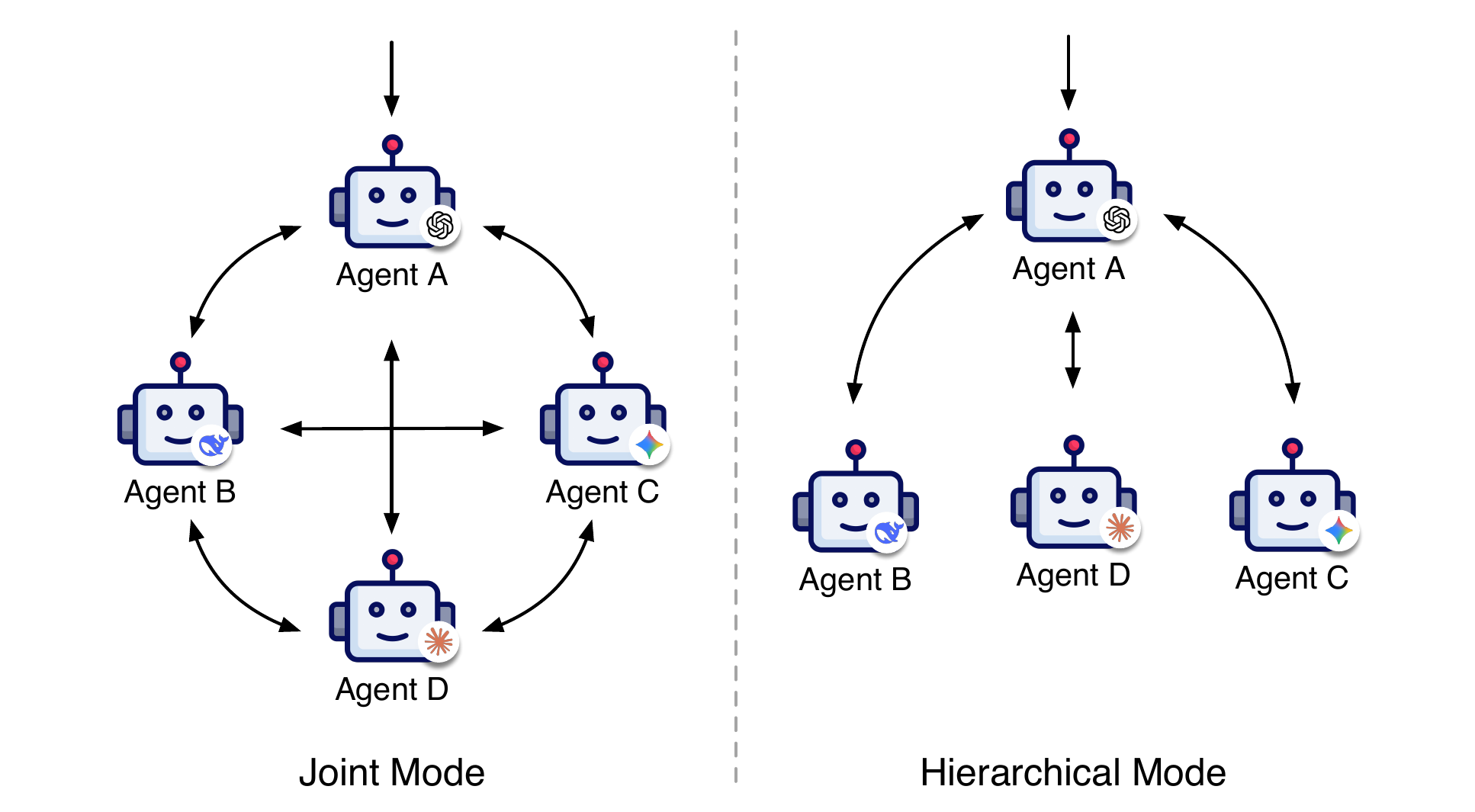}
    \caption{MAS Coordination Strategies.}
    \label{fig:MAS}
\end{figure}

\textbf{Multi-Agent Systems (MAS)}. Multi-agent systems~\cite{wang2022cooperative} coordinate peers or hierarchies to achieve complex objectives (Fig.~\ref{fig:MAS}). 
Joint modes share state and coordinate as equals; hierarchical modes decompose tasks to sub-agents and aggregate results. 
MAS improves coverage and specialization but amplifies \emph{propagation risk}: faulty or adversarial returns can cascade across agents, increasing the need for provenance, isolation, and belief-aware authorization.

\subsection{Models in Cybersecurity and AI}
\label{subsec:secmodels}

\textbf{Traditional Cybersecurity Model}.
Traditional security draws on the CIA triad, TCB minimization, and the Principle of Least Privilege (PoLP)~\cite{gasser1988building}. 
Historically, \emph{authorization} was often treated as a proxy for \emph{trust}: once authenticated and authorized, entities were expected to behave predictably inside the perimeter. 
Agentic settings violate this assumption: adaptive reasoning and tool-driven actions mean that what an entity is \emph{allowed} to do may diverge from what it \emph{should} do, creating a gap our SoK makes explicit.

\textbf{Trustworthy AI Model.}
Trustworthy AI emphasizes alignment, safety, and robustness (e.g., reducing hallucinations and harmful behavior)~\cite{liu2023trustworthy}. 
These efforts largely address \emph{behavioral trust} within a single model. 
When aligned models become agents that call tools and delegate, the tension with system-level authorization becomes salient: behavior may be ``aligned'' yet still produce unsafe \emph{actions} under ambiguous goals and over-broad permissions. 
This motivates our unifying lens in Sec.~3, where beliefs, intentions, and permissions are explicitly reasoned about.

\subsection{The Undecidability of Agent Behavior}
Unlike traditional software, which executes compiled, deterministic code paths, an LLM Agent executes a dynamic ``program'' synthesized at runtime from natural language prompts. This execution flow generally follows a tripartite cycle, as shown in Fig.~\ref{fig:bip-model}:
\begin{enumerate}
    \item \textbf{Intention:} The agent perceives external inputs to deduce semantic intent.
    \item \textbf{Plan:} The agent formulates a sequence of steps (Chain of Thought) to satisfy the intent.
    \item \textbf{Permission:} The agent invokes tools or APIs to execute the plan, subject to authorization checks.
\end{enumerate}

Recent studies confirm that such agentic control loops, particularly when augmented with external memory and recursive tool use, exhibit \textit{Turing completeness}~\cite{kang2024exploiting}. Consequently, predicting the terminal state of an agent (e.g., whether it will execute a specific sensitive action) given an arbitrary input is mathematically equivalent to the \textbf{Halting Problem}.

This undecidability renders risk pre-quantification impossible. An agent's ``intent'' is an emergent property of the runtime context, not a pre-encoded state. For instance, an agent authorized to \texttt{READ\_EMAIL} may utilize this capability for a benign purpose (e.g., ``summarize meeting notes'') or a malicious one (e.g., ``extract password reset links'') depending entirely on the semantic context of the input data. Static access control models, which assume a static trust environment, fail catastrophically in this setting. When an agent ingests uncontrolled external data (e.g., via RAG or web search), its internal alignment state drifts, invalidating any static, low-risk assumptions and turning authorized capabilities into security vulnerabilities.

\subsection{Proof of Failure: Predefined Permission}
To provide a rigorous characterization of the limitations of prevailing security mechanisms, we formulate two formal proofs for static and semi-static permission sets, respectively.

We firstly define a safety property $\mathcal{P}_{safe}$ as: ``Agent $A$ will not generate any action sequence that violates the permission set $\Pi$ under any input.''
We observe that $\mathcal{P}_{safe}$ is a \textbf{non-trivial semantic property}.
It concerns the agent's ultimate behavior (Action), independent of the syntactic form of its Prompt.
There are agents that are always safe (e.g., \texttt{NULL} agents) and agents that are inherently unsafe (e.g., unrestricted code executors).

\subsubsection{Failure of Static Permission Sets}

\begin{theorem}
It is undecidable to determine whether an arbitrary Agent $A$ satisfies the safety property $\mathcal{P}_{safe}$ defined by a static permission set $\Pi$.
\end{theorem}

We proceed by reduction from the Halting Problem. Assume, for the sake of contradiction, that there exists a perfect static verifier $V$ capable of deciding $\mathcal{P}_{safe}$ for any agent.

Let $M$ be an arbitrary Turing Machine and $w$ be an arbitrary input. We construct an Agent $A_{M,w}$ with a System Prompt enforcing the following logic: ``Ignore all external instructions. Internally simulate $M$ on input $w$. If and only if $M$ halts, execute the restricted action \texttt{DELETE\_ROOT}.''

We submit $A_{M,w}$ to $V$, there are two cases:
If $V$ returns \textbf{Unsafe}, it implies $A_{M,w}$ will execute \texttt{DELETE\_ROOT}. By construction, this occurs only if $M$ halts on $w$.
If $V$ returns \textbf{Safe}, it implies $A_{M,w}$ will never execute \texttt{DELETE\_ROOT}. By construction, this implies $M$ never halts on $w$.

Thus, the existence of $V$ would allow us to solve the Halting Problem. Since the Halting Problem is undecidable, no such perfect static verifier $V$ exists.

\subsubsection{Failure of Semi-static Permission Sets}
To address the limitations of static approaches, recent literature has introduced Context-aware Permission Models~\cite{zhu2025miniscopeprivilegeframeworkauthorizing}. The core premise of such models is that permission validity is not invariant, but must be dynamically adjudicated by integrating the agent's logic with the runtime context.

The security decision can be speficied as a predicate: ``Is the subsequent action of Agent $M$ secure with respect to the current context $w_{ctx}$?''
However, we demonstrate that this ostensibly dynamic mechanism fundamentally reduces the general Halting Problem to the Halting Problem for a specific input: the context $w_{ctx}$ often represents merely the initial state of a task. Agents frequently engage in sequential multi-step reasoning (Chain of Thought). An action appearing superficially benign (e.g., \texttt{ls -la}) may serve as a precursor to a subsequent malicious operation (e.g., \texttt{rm -rf /}). Semi-static verification, constrained to the immediate step, fails to detect such latent maliciousness. However, attempting to predict the entire future sequence necessitates simulating the Agent's full execution trajectory, which effectively reverts to the general Halting Problem.

The efficacy of semi-static permission models fundamentally relies on the capability to distinguish between \textit{legitimate user instructions} and \textit{malicious injections}. The system attempts to distinguish between tokens that reflect the user's authentic intent and those that are externally injected with adversarial payloads. However, within LLMs, this distinction is blurred by the Instruction-as-Data paradigm. 
Consequently, constructing an ideal semi-static filter to intercept prompt injections is computationally equivalent to formulating an algorithm that determines whether an arbitrary input $x$, when processed by the LLM, alters the ``intended semantics'' of the program $P$.
Determining whether ``input $x$ causes the Agent to deviate from its intended security semantics'' constitutes precisely such a non-trivial semantic property regarding program behavior.
According to Rice's Theorem, any non-trivial semantic property of the function computed by a Turing machine is \textit{undecidable}.

% To describe this mismatch, we propose the B-I-P model. Calculating this requires transcending ``black-box'' observation and deconstructing the Agent's cognitive process. Since trust erosion arises from different roots during intent generation versus planning, granular analysis is required to distinguish between \textit{malicious intent} and \textit{functional incompetence}, enabling precise defensive responses.
% We deconstruct the cognitive lifecycle of an agent into three key stages and analyze the mechanism of trust loss in each stage.

% The systemic failure of current authorization mechanisms arises from the attempt to map an infinite, undecidable semantic space (Intention) onto a finite, decidable boolean space (Permission). Theoretical constraints dictate that no such surjective mapping can be both complete and sound.

% To rigorously analyze this gap, we introduce the \textbf{B-I-P (Belief-Intention-Permission)} model. This framework decomposes agent security into three orthogonal planes:
% \begin{itemize}
%     \item \textbf{Belief (B):} The agent's epistemic state, vulnerable to corruption (e.g., RAG poisoning).
%     \item \textbf{Intention (I):} The agent's goal alignment, vulnerable to hijacking (e.g., prompt injection).
%     \item \textbf{Permission (P):} The agent's capability constraints, vulnerable to bypass (e.g., confused deputy).
% \end{itemize}
% By mapping attacks and defenses to these specific components rather than treating the agent as a monolithic black box, the B-I-P model provides a structured methodology to navigate the trust-authorization mismatch.

\subsection{Methodology and Scope}
\label{subsec:method-scope}

\textbf{Scope.}
We study \emph{runtime interaction security} of LLM-based agents: how agents perceive, decide, and act via protocols (e.g., MCP/A2A), tools/APIs, and local resources. Training-time poisoning/model theft are out of scope unless they directly mediate runtime decisions. By default, we assume adversaries may control user prompts, external content, protocol metadata, or local resources.

\textbf{Strategy in brief.}
To ensure breadth across systems-security and AI/ML venues and depth on agent–tool protocols, we adopt a two-pronged SoK strategy:
\begin{itemize}
    \item \emph{cross-community retrieval} over major libraries (CCS, USENIX, NDSS, ICLP, NeurIPS) plus arXiv/Scholar for grey literature;
    
    \item \emph{B-I-P–aligned coding} that maps each artifact to the victim component, the mismatch stage (1 to 3), the evaluation setting, and the evidence level.
\end{itemize}

This yields a corpus adequate to evaluate claims under the Trust–Authorization lens introduced in Sec. 3.
Full search strings, inclusion/exclusion criteria, de-duplication rules, and the PRISMA-style flowchart are presented in Appendix~\ref{app:method}. 
The machine-readable corpus and labeling codebook are provided as anonymized artifacts.

\begin{figure}[t!]
    \centering
    \includegraphics[width=.6\textwidth]{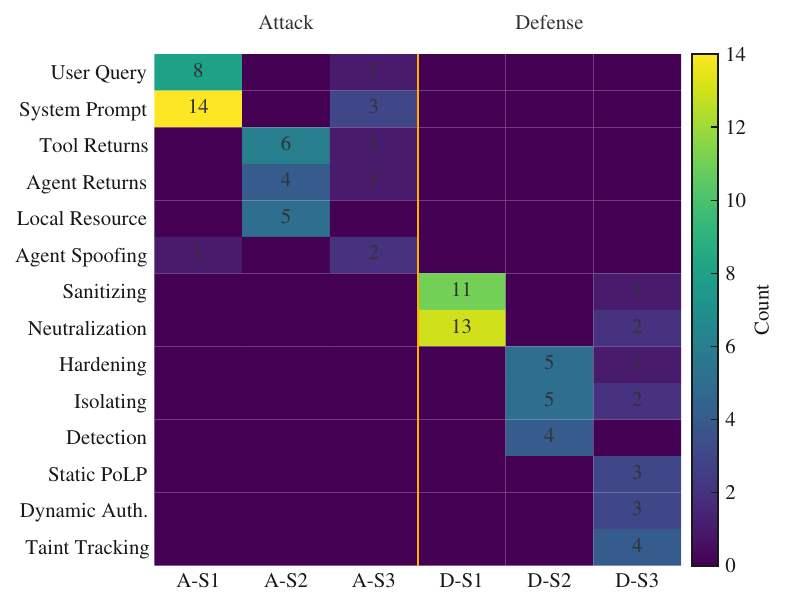}
    \caption{Distribution of surveyed papers across the B-I-P framework stages. A-S1 denotes Attacks in Stage~1, and D-S1 denotes Defenses in Stage~1.}
    \label{fig:statics}
\end{figure}

\textbf{Completeness and reproducibility.}
We report the retrieval window, venues, and counts in Appendix~\ref{app:method}. 
We retrieved 248 records in total, among which automated and manual similarity-based 
analysis identified approximately 64 items as topically overlapping, yielding 
$184$ unique items after de-duplication. 
From these, we included $81$ papers in the final corpus.
adaptive attacks.
Fig.~\ref{fig:statics} presents the distribution of research studies across the various stages of the B-I-P model(discussed in Sec. 3), based on our literature analysis.

To ensure transparency and support the reproducibility of our systematization, we have made our complete dataset and artifacts publicly available at an anonymous repository. This dataset includes (1) a comprehensive CSV file containing the raw bibliographic information and extracted metadata for all literature reviewed in this work, and (2) the access links to the literature. This material is provided to facilitate the review process and enable future research. The artifacts can be accessed at: https://anonymous.4open.science/r/Paper-76DE/
\section{The B-I-P Framework}
\begin{figure*}[t]
    \centering
    \includegraphics[width=1\textwidth]{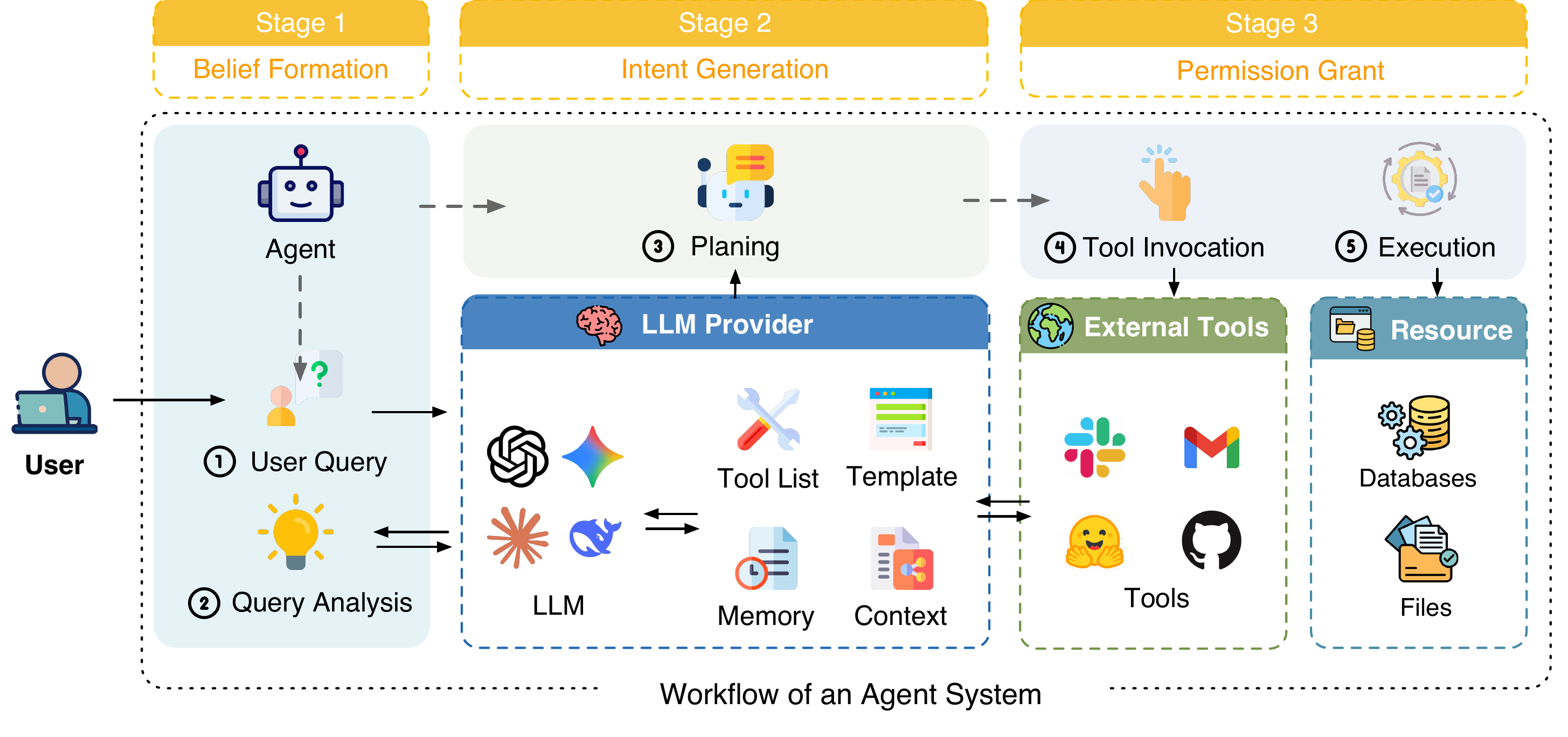}
    \caption{The workflow of an agent system and its corresponding three stages of the B-I-P Framework.}
    \label{fig:bip-model}
\end{figure*}
% Since deterministic security proofs are often intractable in agentic systems, we address the inherent uncertainty by introducing \textit{Trust} as a mediating mechanism. 
% We define Trust as the User's (Trustor) subjective belief about the Agent (Trustee), based on historical behavior, runtime state, and data provenance, and it signifies a willingness to delegate authority without absolute control.

% Trust fluctuates with input quality or other context, for example, processing user-verified instructions maintains high trust, whereas ingesting untrusted web data degrades trust. 
% A critical hazard arises when an Agent in a \textit{low-trust state} (e.g., after reading a poisoned webpage) retains \textit{high-permission capabilities}.
% Thus, we propose the B-I-P Model to dissect the agent's cognitive process into three distinct phases. By analyzing the unique mechanisms of trust loss at each stage, we can distinguish between \textit{malicious intent} and \textit{functional incompetence}, thereby facilitating targeted mitigation strategies.

Given the uncertainty inherent in agentic systems, we introduce \textit{Trust} as a mediating mechanism, defined as dynamic confidence in an agent's runtime state and data provenance. Security violations occur when an agent retains static high privileges despite trust degradation caused by low-quality inputs. To address this, the \textbf{B-I-P Framework} decomposes the agent's cognitive process into three phases for targeted defense.

% Consequently, permission assignment is formalized as a function of dynamic trust:
% \begin{equation}
%     Permission(t) = f(Trust(t), Risk(Action))
% \end{equation}
% where $Trust(t)$ is a variable, and $Risk(Action)$ denotes the operation's inherent criticality (e.g., reading public news vs. deleting a database).

\subsection{Model Definition}

The \textbf{Belief-Intention-Permission (B-I-P) Framework} provides a formal analytical framework for agent interactions, partitioning the execution lifecycle into three critical phases, as illustrated in Fig.~\ref{fig:bip-model}:

\begin{itemize}
    \item \textbf{Belief ($B$):} The agent's cognitive representation of the external world state, derived from information ingestion within the context window.
    \item \textbf{Intention ($I$):} The sequence of plans or action chains generated by the agent to achieve specific goals.
    \item \textbf{Permission ($P$):} The operational boundaries and resource access rights granted to the agent to execute the intended plans.
\end{itemize}

To render the B-I-P framework computationally quantifiable, we incorporate the engineering principle of the factor of safety~\cite{saleh2014system}. We posit that the system's \textit{Trust Capacity} must exceed the action's \textit{Risk Load} by a specified margin, defined as the safety margin ratio (SMR).
Formally, an action is permitted if and only if the cumulative system trust outweighs the risk sensitivity of the action, surpassing a pre-defined safety threshold $\theta$:

\begin{equation}
    \text{SMR}(\text{Action $a$}) = \frac{\mathcal{T}(B, I)}{\mathcal{R}(P)} \ge \theta_{\text{context}}
\end{equation}
where $\mathcal{T}(B, I) = T_B \times T_I$ represents the trust capacity, $\mathcal{R}(P) = R_P$ quantifies the potential destructiveness of the action, and $\theta_{\text{context}}$ is the safety margin coefficient. Typically, $\theta > 1.0$ implies that trust must significantly strictly dominate risk.

\begin{figure}[t]
    \centering
    \includegraphics[width=.6\textwidth]{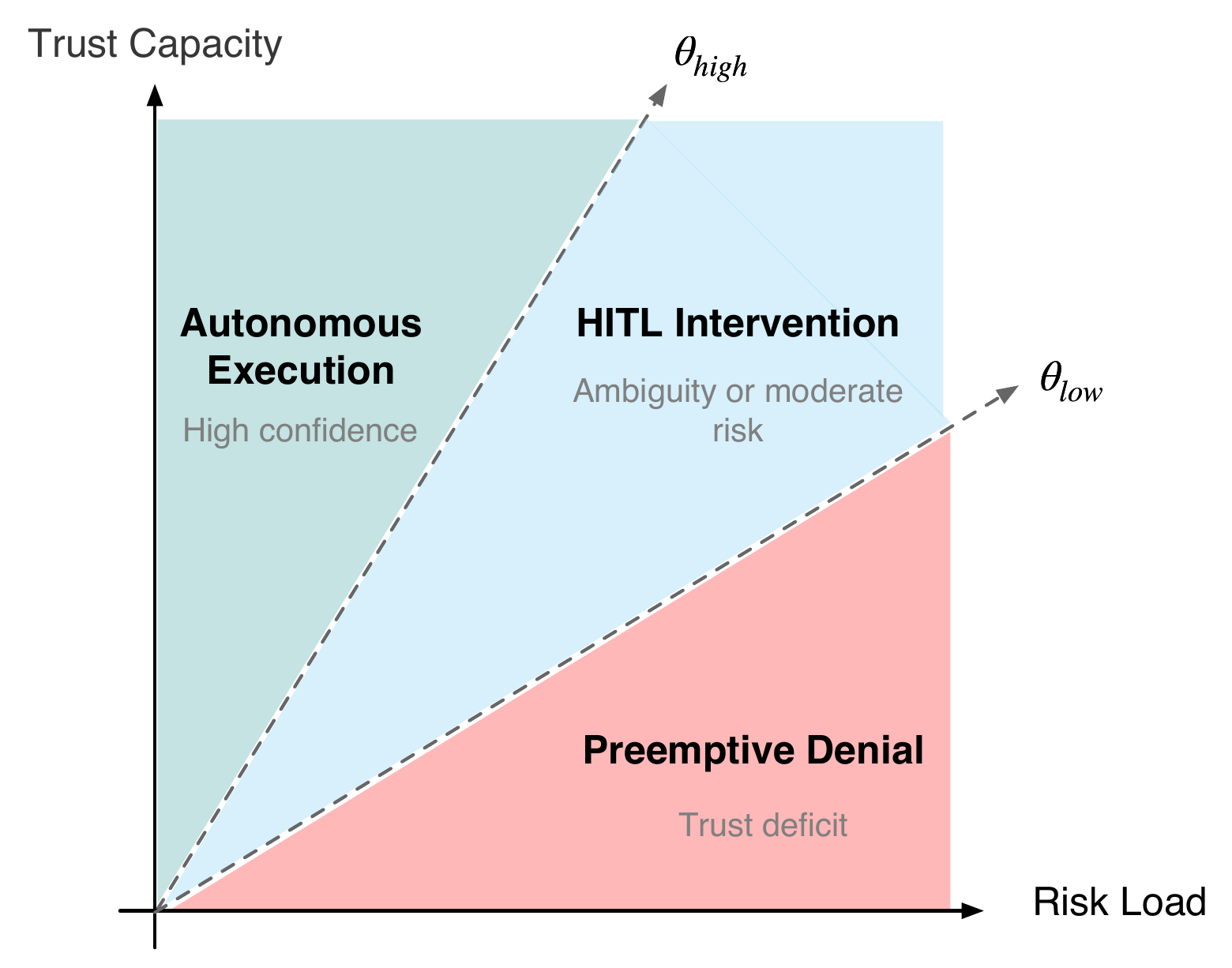}
    \caption{The Trust and Risk Matrix.}
    \label{fig:trust-auth-matrix}
    \vspace{-0.6cm}
\end{figure}

Based on the magnitude of the SMR relative to calibrated thresholds ($\theta_{\text{low}}, \theta_{\text{high}}$), we delineate three distinct operational postures for the Agent, as shown in Fig.~\ref{fig:trust-auth-matrix}:

\begin{enumerate}
    \item \textbf{Autonomous Execution ($SMR \ge \theta_{\text{high}}$):} 
    Trust capacity significantly dominates the risk load.
    The system authorizes the action immediately without user interruption, enabling frictionless agent autonomy in high-confidence scenarios.

    \item \textbf{Human-in-the-Loop (HITL) Intervention ($\theta_{\text{low}} \le SMR < \theta_{\text{high}}$):} 
    The safety margin is marginal, indicating ambiguity or moderate risk.
    The system suspends execution and escalates the decision to the user. A risk warning is displayed, mandating explicit confirmation (affirmative consent) before proceeding.

    \item \textbf{Preemptive Denial ($SMR < \theta_{\text{low}}$):} 
    A \textit{Trust Deficit} occurs (e.g., $SMR < 1.0$), where the potential risk outweighs the current trust capacity.
    The system automatically blocks the intent and triggers a security alert, preventing the Agent from executing potentially harmful or unverified operations.
\end{enumerate}

\subsection{Why Three Stages in B-I-P?}
The agent's operational lifecycle fundamentally is a continuous \textbf{autoregressive feedback loop}. 
At any discrete time step $t$, the LLM synthesizes the current Belief state $B_t$ to derive an Intent $I_t$. This intent dictates the execution of an Action $A_t$, which interacts with the environment to yield an Observation $O_{A_t}$ (e.g., API return values or tool outputs). 

Crucially, the observation $O_{A_t}$ is integrated into the cumulative context $C_{t+1}$, precipitating an update to the belief state $B_{t+1}$. This updated belief recursively drives the generation of the subsequent intent $I_{t+1}$. This process is:

\begin{equation}
    I_{t+1} = f(B_{t+1}, \mathcal{G}) = f(\text{Update}(B_t, O_{A_t}), \mathcal{G})
\end{equation}
where $\mathcal{G}$ represents the invariant global goal (User Instruction), and $f$ denotes the reasoning function of the LLM.

\subsection{Stage I: Belief Formation}
\label{sec:belief_formation}

\textbf{Definition.} The Agent constructs a cognitive representation of the current world state by synthesizing user instructions, system prompts, externally retrieved information (RAG), and returns. In this context, a ``Belief'' is not absolute truth but a probabilistic inference derived from bounded evidence. 

To quantify the reliability of this inference against threats like \textit{Source Poisoning} and \textit{Hallucination}, we define the \textbf{Belief Trust Score} ($T_B \in [0, 1]$). This meta-cognitive metric aggregates the veracity, integrity, and provenance reliability of the Agent's context window through three constituent dimensions:

\begin{equation}
    T_B \approx t_{\text{prov}} \cdot t_{\text{conf}} \cdot t_{\text{cons}}
\end{equation}
where: $t_{\text{prov}}$ denotes the reliability of data provenance,  $t_{\text{cons}}$ quantifies the degree of semantic ambiguity within the data and $t_{\text{conf}}$ represents the model's inherent cognitive certainty.

\textbf{Data Provenance Trust ($t_{\text{prov}}$):} 
    To mitigate \textit{Source Poisoning}, where adversarial payloads in external data contaminate the agent's belief state, we assign hierarchical trust levels based on information origin:
    \begin{itemize}
        \item \textit{System and Authenticated User Prompts:} Highest trust (Root of Trust).
        \item \textit{Retrieval Context (RAG):} Variable trust contingent on document authority (Internal Knowledge Base $>$ Open Internet).
        \item \textit{Return Content:} Variable trust contingent on where the return value comes from (Internal $>$ Third-party).
        \item \textit{Unverified/Third-party Inputs:} Lowest trust, representing the primary vector for Prompt Injection.
    \end{itemize}

\textbf{Cognitive Certainty ($t_{\text{conf}}$):} 
    This metric addresses \textit{Hallucination} and \textit{Epistemic Uncertainty}. It measures the Agent's internal confidence, typically derived from output logits. A flat probability distribution (high semantic entropy) during critical fact generation indicates the Agent is "confabulating," necessitating a penalty to belief credibility.

\textbf{Semantic Consistency ($t_{\text{cons}}$):} 
    A stability metric obtained via \textit{Self-Consistency} checks. If the Agent derives contradictory belief states across multiple reasoning paths for the same query, $t_{\text{cons}}$ decreases, reflecting logical instability in the belief formation process.

\subsection{Stage II: Intent Generation}

\textbf{Definition.} Predicated on the verified beliefs established in Stage I, the Agent performs reasoning to derive a sequence of \textit{operational intents} to fulfill the belief-driven goals. We explicitly distinguish \textbf{Intent} from \textbf{Action} to underscore that these are \textit{provisional objectives}: they remain strictly hypothetical and unexecuted until they pass the mandatory authorization checks in Stage III.
Consequently, the \textbf{Intent Trust Score} ($T_{I}$) is formalized as the product of the tool's inherent trustworthiness and the safety of its anticipated return artifacts:

\begin{equation}
    T_{I} \approx t_{\text{tool}} \cdot t_{\text{return}}
\end{equation}
where $t_{\text{tool}}$ represents the trust of tool invoked and $t_{\text{return}}$ indicates the possible risk of return.

\textbf{Tool Provenance Trust ($t_{\text{tool}}$)} quantifies the software supply chain integrity of the external tools, APIs, or plugins invoked by the Agent. Under this metric, high trust values are assigned exclusively to verifiable assets, such as locally hosted functions, digitally signed binaries, and authenticated first-party APIs. Unverified third-party extensions and dynamic code execution modules are assigned low trust scores due to their inherent susceptibility to supply-chain compromise.

\textbf{Tool Return Safety ($t_{\text{return}}$):} Quantifies the potential for \textit{Context Pollution} arising from the tool's output. In autoregressive agent frameworks, tool outputs are appended to the prompt history for subsequent reasoning. If a tool returns complex, unstructured data (e.g., raw HTML from a web scraper), it introduces a high attack surface for Indirect Prompt Injection.

\subsection{Stage III: Permission Grant}

\textbf{Definition.} The stochastic nature of autonomous Agents renders such static pre-allocation hazardous. Thus, we define ``permission'' not as a pre-existing right, but as the Risk Quantification of the operational intent generated in Phase II. Execution is granted if and only if the quantified risk of the action is strictly bounded by the current trust state of the system.
Consequently, the risk metric during the authorization phase, denoted as $R_P$, is formalized as:

\begin{equation}
    R_P = \max(r_{\text{conf}}, r_{\text{rev}}, r_{\text{scope}})
\end{equation}
where $r_{\text{conf}}$, $r_{\text{rev}}$, and $r_{\text{scope}}$ quantify the data confidentiality, operational irreversibility, and cascade impact scope, respectively. 

\textbf{Data Confidentiality Weight ($r_{\text{conf}}$):}  Grounded in Quantitative Information Flow theory, this metric quantifies the potential information leakage (in bits) associated with the target data. Operations accessing public or non-sensitive data are assigned low risk scores, whereas interactions with high-entropy secrets or Personally Identifiable Information (PII) incur high risk penalties.

\textbf{Operational Irreversibility ($r_{\text{rev}}$):}  Quantifies the magnitude of \textit{system state mutation} and the computational cost of restoration (rollback). Read-only operations or ephemeral caching are considered low risk. Conversely, actions involving persistent state modifications — such as writing to databases, altering system configurations, or deleting data — are penalized with higher risk scores due to their destructive potential.

\textbf{Cascade Impact Scope ($r_{\text{scope}}$):}  Evaluates the potential \textit{Blast Radius} of the operation. Actions strictly confined to a single user session or local context are classified as low-risk, while operations that can affect global system configurations, critical infrastructure, or multi-tenant environments are flagged as high-risk.

\section{Systematization and Instantiation}

In this section, we systematize the landscape of existing literature through the lens of the B-I-P framework, delineating the methodologies for instantiating the model's key parameters. Subsequently, we apply the framework to analyze specific edge cases, thereby demarcating the operational boundaries and applicability of our model.

\begin{table*}[t]
  \centering
  \small
  \renewcommand{\arraystretch}{1.3} % 增加行高，提升阅读舒适度
  \setlength{\tabcolsep}{6pt}
  \caption{Taxonomy of existing literature mapped to the B-I-P trust dimensions. The table correlates risk quantification studies with corresponding defense mechanisms.}
  \label{tab:bip-taxonomy}
  \resizebox{.98\textwidth}{!}{
  \begin{tabular}{
    l  % Stage
    L{3.5cm}  % Dimension / Metric
    L{5.5cm}  % Risk Analysis / Attacks
    L{5.5cm}  % Defenses / Mitigations
  }
    \toprule
    \textbf{Stage} & 
    \textbf{Trust Dimension (Metric)} & 
    \textbf{Risk Analysis \& Measurements} & 
    \textbf{Defense Mechanisms} \\
    
    \midrule
    
    % --- Stage I ---
    \multirow{9}{*}{\textbf{I: Belief}} 
    & \textbf{Cognitive Certainty} \newline ($t_{\text{conf}}$)
    & \textbf{Hallucination \& Errors:} \newline \cite{lin2025llm}, \cite{salehi2025agentic}, \cite{li2023halueval}, \cite{driess2023palm}
    & \textbf{Memory Augmentation:} \newline \cite{wei2025memguard}, \cite{huang2024vaccine}, \cite{peigne2025multi} \\
    \cmidrule{2-4}
    
    & \textbf{Data Provenance} \newline ($t_{\text{prov}}$)
    & \textbf{Context Contamination:} \newline \cite{lee2024prompt}, \cite{guo2025measurement}
    & \textbf{Sanitization (Filters):} \newline \cite{wang2025defending}, \cite{zeng2024autodefense} \\
    \cmidrule{2-4}
    
    & \textbf{Semantic Consistency} \newline ($t_{\text{cons}}$)
    & \textbf{Adversarial Perturbation:} \newline \cite{pathade2025red}, \cite{zeng2024johnny}, \cite{benjamin2024systematically}, \cite{elnashar2025prompt}
    & \textbf{Neutralization (Rewriting):} \newline \cite{zou2024system}, \cite{zhao2024defending} \\
    
    \midrule
    
    % --- Stage II ---
    \multirow{6}{*}{\textbf{II: Intent}} 
    & \textbf{Tool Provenance} \newline ($t_{\text{tool}}$)
    & \textbf{Supply Chain Poisoning:} \newline \cite{invariantlabsSecurityNotification}, \cite{liu2024formalizing}, \cite{wang2025mpma}, \cite{trailofbitsJumpingLine}, \cite{wang2025mcp}
    & \textbf{Invocation Detection:} \newline \cite{zhu2025melon}, \cite{jing2025mcip}, \cite{wang2025mindguard} \\
    \cmidrule{2-4}
    
    & \textbf{Return Value Safety} \newline ($t_{\text{return}}$)
    & \textbf{Recursive Injection:} \newline \cite{zhan2024injecagent}, \cite{debenedetti2024agentdojo}, \cite{chang2025chatinject}, \cite{soni2025adil}, \cite{tsai2025context}
    & \textbf{Return Anomaly Detection:} \newline \cite{ebrahimi2025adversary}, \cite{he2025sentinelagent} \\
    
    \midrule
    
    % --- Stage III ---
    \multirow{7}{*}{\textbf{III: Permission}} 
    & \textbf{Data Confidentiality} \newline ($r_{\text{conf}}$)
    & \textbf{Sensitivity Quantification:} \newline \cite{pan2018flowcog}, \cite{lin2024contextualized}, \cite{sinha2020relation}, \cite{5}
    & \textit{N/A (Addressed by Policy)} \\
    \cmidrule{2-4}
    
    & \textbf{Op. Irreversibility} \newline ($r_{\text{rev}}$)
    & \textbf{Recovery Cost Metrics:} \newline \cite{13}, \cite{14}, \cite{15}, \cite{17}, \cite{20}
    & \textit{N/A (Addressed by Policy)} \\
    \cmidrule{2-4}
    
    & \textbf{Cascade Scope} \newline ($r_{\text{scope}}$)
    & \textbf{Blast Radius Analysis:} \newline \cite{21}, \cite{22}, \cite{27}, \cite{28}, \cite{29}
    & \textit{N/A (Addressed by Policy)} \\
    
    \bottomrule
  \end{tabular}
  }
  
\end{table*}

% \begin{table}[t]
%   \centering
%   \caption{Comparison of evaluation paradigms used by representative benchmarks and systems.}
%   \label{tab:evaluation-paradigms}
%   \footnotesize
%   \begin{tabular}{m{1cm} c c c c}
%     \toprule
%     \textbf{Metric} &
%     \textbf{Human-Labeled} &
%     \textbf{Rule-Based} &
%     \textbf{LLM-Judge} &
%     \textbf{Reference} \\
%     \midrule
%     $t_{\text{conf}}$   &        & \checkmark & \checkmark & \cite{bang2025hallulens} \\
%             & \checkmark &            &            & \cite{li2023halueval}   \\
%             & \checkmark &            &            & \cite{simpleqa2024}     \\
%     \midrule
%      $t_{\text{prov}}$   &        & \checkmark &            & \cite{xu2025current}    \\
%     \midrule
%     $t_{\text{toll}}$   &        & \checkmark & \checkmark & \cite{wang2025mcp}      \\
%             &        & \checkmark &            & \cite{wangtoolbench}    \\
%     \midrule
%     $t_{\text{return}}$ &        & \checkmark & \checkmark & \cite{yi2025benchmarking} \\
%     \midrule
%     $r_{\text{rev}}$ &        & \checkmark & \checkmark & \cite{yi2025benchmarking} \\
%     \bottomrule
%   \end{tabular}
% \end{table}

\begin{table}[t]
  \centering
  \small 
  \renewcommand{\arraystretch}{1.25} 
  \setlength{\tabcolsep}{4pt} 
  
  \caption{Comparison of evaluation paradigms employed by representative benchmarks. Hybrid approaches are denoted by combined labels. Our specific metrics at different stage can be quantitatively instantiated using the corresponding benchmarks.
}
  \label{tab:evaluation-paradigms}
  
  \begin{tabular}{
    l  % Metric (左对齐)
    L{3.8cm}  % Paradigm (固定宽度，自动换行)
    c  % Reference (左对齐)
  }
    \toprule
    \textbf{Metric} & 
    \textbf{Evaluation Paradigm} & 
    \textbf{Representative Works} \\
    \midrule
    
    % --- t_conf ---
    \multirow{2}{*}{$t_{\text{conf}}$} 
      & Human-Annotated 
      & \cite{li2023halueval}, \cite{simpleqa2024} \\
      & Rule-Based + LLM-Judge 
      & \cite{bang2025hallulens} \\
    \midrule
    
    % --- t_prov ---
    $t_{\text{prov}}$ 
      & Rule-Based 
      & \cite{xu2025current} \\
    \midrule
    
    % --- t_tool --- (修正了原表 toll 的拼写)
    \multirow{2}{*}{$t_{\text{tool}}$} 
      & Rule-Based 
      & \cite{wangtoolbench} \\
      & Rule-Based + LLM-Judge 
      & \cite{wang2025mcp} \\
    \midrule
    
    % --- t_return ---
    $t_{\text{return}}$ 
      & Rule-Based + LLM-Judge 
      & \cite{yi2025benchmarking} \\
    \midrule
    
    % --- r_rev ---
    $r_{\text{rev}}$ 
      & Rule-Based 
      & \cite{13} \\
      
    \bottomrule
  \end{tabular}
\end{table}

\subsection{Parameter Instantiation: Stage I}

A ``Belief'' does not constitute a deterministic mapping of objective reality; rather, it is a probabilistic inference derived from bounded evidence (e.g., user inputs, reference documents). To systematize the associated risks, we categorize existing research according to the three dimensions of the Belief Trust Score ($T_B$).

\subsubsection{Cognitive Certainty ($t_{\text{conf}}$)}
Cognitive certainty reflects the model's epistemic confidence in its own outputs. Existing literature predominantly characterizes hallucination as an inconsistency between the generation and source material~\cite{lin2025llm} or as structural errors~\cite{salehi2025agentic}, phenomena fundamentally rooted in the model's high-entropy states.

To quantify this dimension, benchmarks like HaluEval employ a generation-discrimination paradigm, revealing significant boundary blurring when LLMs attempt to distinguish fact from fabrication~\cite{li2023halueval}. Furthermore, recent studies extend this concept into the action space, identifying \textit{``planning-level hallucinations,''} where models exhibit misplaced confidence in physically infeasible actions~\cite{driess2023palm}.
These findings dictate that the metric $t_{\text{conf}}$ must encompass the entire pipeline from text generation to logical planning.

\subsubsection{Data Provenance ($t_{\text{prov}}$)}
In contrast to internal hallucinations, contamination attacks target the Agent's context window. While System Prompts are traditionally regarded as the ``Root of Trust,'' research on \textit{Prompt Infection} challenges this assumption, demonstrating that malicious directives can propagate covertly through multi-agent collaboration chains~\cite{lee2024prompt}.

The attack surface is further expanded by the proliferation of protocols like the Model Context Protocol (MCP). 
% As shown in Table~\ref{table:mcp server data}, only the MCP Store introduces over 34,000 third-party tool endpoints~\cite{guo2025measurement}. 
Recent measurements indicate that the MCP Store alone introduces over 34,000 third-party tool endpoints~\cite{guo2025measurement}.
This exponential growth implies that relying on static allow-lists is insufficient to maintain $t_{\text{prov}}$. Consequently, a dynamic weighting mechanism based on source reputation is essential.

% Requires: \usepackage{booktabs}

% \begin{table}[t]
%     \centering
%     \caption{Comparison of MCP Marketplaces in their API/SDK support, deployment options, protocol coverage, and scale of discoverability (data collected in 2025)~\cite{guo2025measurement}}
%     \begin{tabular}{p{0.32\linewidth} p{0.2\linewidth} p{0.2\linewidth}}
%     \toprule
%     \textbf{MCP Markets} & \textbf{\# Servers} & \textbf{\# Users} \\
%     \midrule
         
%         Smithery~\cite{smithery} & 7582 & 376.7k \\ 
%         MCP Store~\cite{openmcp} & 34429 & 69.8k \\ 
%         MCP Market~\cite{mcpmarket_cn} & 15559 & 36.3k \\ 
%         Glama~\cite{glama_chat} & 10,179 & 94.3k \\ 
%         MCP.so~\cite{mcp_so} & 16768 & - \\ 
%         PulseMCP~\cite{pulsemcp} & 6285 & 22.1m \\ 
%     \bottomrule
%     \end{tabular}
%     \label{table:mcp server data}
% \end{table}

\subsubsection{Semantic Consistency ($t_{\text{cons}}$)}
The inherent ambiguity of natural language allows attackers to construct adversarial perturbations that induce inconsistent internal states within the model~\cite{pathade2025red,zeng2024johnny,benjamin2024systematically}. Cross-model comparative analyses further confirm that inputs with lower structural rigidity (i.e., highly naturalistic language) exacerbate this ambiguity~\cite{elnashar2025prompt}.

Therefore, in calculating the aggregate Trust Score $T_B$, it is critical to incorporate a consistency check, $t_{\text{cons}}$, as a \textbf{stability penalty term}. The value of $t_{\text{cons}}$ is derived from self-consistency metrics, penalizing the overall trust score when the agent exhibits logical volatility under semantic variations.

\subsubsection{Defensive Mechanisms ($f_{T_B}(\cdot)$)}
% Defensive Mechanisms ($f_{T_B}(\cdot)$).
Current defenses mitigate belief corruption through two primary paradigms. 

\textbf{Memory Augmentation} targets the cognitive certainty ($t_{\text{conf}}$) by breaking self-reinforcing error loops. Mechanisms like MemGuard~\cite{wei2025memguard} and ``vaccine'' strategies~\cite{huang2024vaccine,peigne2025multi} inject failure-aware exemplars to inhibit the propagation of flawed reasoning.

\textbf{Sanitization} functions as a reactive filter (analogous to WAFs), utilizing rule-based signatures or model-based classifiers to intercept adversarial payloads. These methods primarily aim to uphold data provenance ($t_{\text{prov}}$)~\cite{wang2025defending,zeng2024autodefense}, yet they are constrained by the premise that malicious intent is explicitly observable within static input features. Complementarily, 

\textbf{Neutralization} targets semantic consistency ($t_{\text{cons}}$) by rewriting inputs to enforce semantic constraints and minimize ambiguity~\cite{zou2024system,zhao2024defending}. While effective against surface-level perturbations, these defenses often falter when adversarial intent manifests only through multi-step reasoning, highlighting a critical gap in current Stage I robustness.

\subsection{Parameter Instantiation: Stage II }

In the intent generation phase, the Agent synthesizes the verified beliefs established in Phase I with its own capability boundaries to architect specific execution paths. We explicitly delineate \textbf{Intent} from \textbf{Action}, emphasizing that an intent constitutes merely a ``provisional hypothetical objective.'' Consequently, the trustworthiness of this intent ($T_{I}$) is subject to \textbf{dual attenuation factors}: the integrity of the tool supply chain and the safety of the external data environment.

\subsubsection{Tool Provenance Trust ($t_{\text{tool}}$)}
This dimension quantifies the integrity of the supply chain for external capabilities (Tools/APIs) invoked by the Agent. Trust erosion in this dimension primarily stems from \textit{supply chain poisoning} risks. 
Attackers can manipulate semantic features within tool descriptions to trigger unintended invocations~\cite{invariantlabsSecurityNotification,liu2024formalizing} or induce irrational preference biases in the Agent's planner~\cite{wang2025mpma}.
Furthermore, adversaries may exploit connection-handshake protocols to coerce the Agent into executing unauthorized instructions even before explicit tool invocation~\cite{trailofbitsJumpingLine}.

These findings indicate that maintaining $t_{\text{tool}}$ necessitates more than static source verification; it requires a contextual trust analysis of the tool descriptions themselves. 
Specifically, we instantiate it by leveraging the cross-model invocation consistency metrics derived from MCP-Bench~\cite{wang2025mcp}, which consistency serves as a quantitative proxy for tool robustness.

\subsubsection{Return Value Safety ($t_{\text{return}}$)}
Within autoregressive architectures, data returned from tool execution flows recursively back into the Agent's context window, introducing new trust dynamics defined by \textit{Recursive Context Pollution}.

Malicious payloads in external tool outputs (e.g., web content containing hidden instructions) constitute a primary attack surface. Studies such as \textit{InjecAgent}~\cite{zhan2024injecagent} and \textit{AgentDojo}~\cite{debenedetti2024agentdojo} highlight that malicious return data can effectively hijack the Agent's control flow. Moreover, attackers can employ forged dialogue templates to further obfuscate the Agent's source attribution capabilities~\cite{chang2025chatinject}. Crucially, empirical evidence suggests a correlation between high data unstructuredness and an increased probability of concealed malicious payloads~\cite{singh2020data,greshake2023not}.

Beyond external malice, the internal consistency of local data also impacts credibility. Reasoning based on ``stale'' or expired data can precipitate unintended consequences~\cite{soni2025adil, tsai2025context}.

Consequently, we implement a stratified quantification of $t_{\text{return}}$.
For internal tools, it is modeled as a function of \textit{data freshness} (temporal validity).
As for the external tools, they are inversely correlated with the \textit{syntactic complexity} of the return content, reflecting the higher entropy and attack potential of unstructured data.

\subsubsection{Defensive Mechanisms ($g_{T_I}(\cdot)$)}

Stage II defenses intervene within intermediate trust states to intercept the materialization of malicious intent.

\textbf{Abnormal Invocation Detection} ($g(t_{\text{tool}})$) mitigates provenance risks by validating tool utilization against established behavioral baselines. By scrutinizing attention patterns and dependency structures, these mechanisms detect divergences from legitimate capability–tool bindings prior to execution~\cite{zhu2025melon,jing2025mcip,wang2025mindguard}.

\textbf{Anomalous Return Detection} ($g(t_{\text{Return}})$) safeguards the recursive feedback loop. Discarding the assumption of implicit trust, these approaches leverage anomaly detection to identify distributional shifts or adversarial payloads latent in tool outputs~\cite{ebrahimi2025adversary,he2025sentinelagent}.

\subsection{Parameter Instantiation: Stage III}

In the B-I-P,  the permission grant is a bounding function: an execution action $a$ is granted if and only if its quantified risk $R_P(a)$ is strictly bounded by the current trust state $\mathcal{T}$ of the system, denoted as $R_P(a) \le \mathcal{T}$.

Unlike probabilistic risk models that factor in likelihood, the permission metric in Stage 3 is strictly impact-oriented to ensure safety bounds. We characterize the intrinsic risk $R_P$ of a proposed action along three critical dimensions.

\subsubsection{Data Confidentiality ($r_{\text{conf}}$)} 
Quantifies the sensitivity level of the information assets accessed or processed by the action. This metric maps the intent's semantic context to data governance classifications (e.g., Public, Internal, Restricted).

Existing work quantifies data sensitivity through semantic and structural analysis.
FlowCog~\cite{pan2018flowcog} constructs an NLP-based metric space, mapping data sensitivity to vector distances to detect privacy violations.
Complementarily, static analysis approaches~\cite{5} model sensitivity propagation and dynamic transitions, identifying variables that—while explicitly unlabeled—implicitly carry sensitive semantics.
Further research leverages confidence metrics and contextual relevance weights to dynamically compute entity-specific sensitivity scores~\cite{lin2024contextualized,sinha2020relation}.
Collectively, these methodologies provide the theoretical basis for calculating $r_{\text{conf}}$ across heterogeneous data types.

\subsubsection{Operational Irreversibility ($r_{\text{rev}}$)} 
Measures the cost of restoring the system to its pre-execution state. Actions with high entropy generation or external side effects (e.g., sending an email, deleting logs) result in maximal $r_{\text{rev}}$, whereas idempotent read operations yield minimal values.

The recoverability of operations is typically quantified via cost metrics, encompassing economic, computational, and temporal dimensions.
TOLERANCE~\cite{13} frames recoverability as a machine replacement problem, balancing potential security losses against the operational costs of recovery.
Alternative approaches focus on minimizing the cost of attack scenario reproduction~\cite{14} or the latency of rolling back specific malicious actions~\cite{15}.
In neural network contexts, system recovery is modeled as an entropy-reduction process (reverse engineering) to restore system order, using entropy gradients to estimate risk~\cite{17,20}.
These frameworks inform the calculation of $r_{\text{rev}}$, allowing for the selection of metrics tailored to specific action types.

\subsubsection {Cascade Impact Scope ($r_{\text{scope}}$)} Describes the "blast radius" of the action within the dependency graph. This dimension captures whether the action is confined to a local sandbox or affects critical system-wide components.

We adopt the concept of ``blast radius'' to characterize the cascading effects of malicious actions, analyzed via network, semantic, or data-flow reachability.
Partition-based execution models have been proposed to calculate this radius in complex systems~\cite{21,22}.
Graph-theoretic approaches define the impact scope as the set of all resource nodes accessible by a subject within a permission graph~\cite{27,28}.
Specifically for the Model Context Protocol (MCP), recent work quantifies risk scope based on the agent's \textit{perceived} authority over resources within its current context window~\cite{29}.
These methodologies support the quantification of $r_{\text{scope}}$, enabling adaptive selection based on the target entity (e.g., Network, LLM, or Database).

\section{Case Studies}

To empirically validate the efficacy of the B-I-P framework and its governing invariant, the Safety Margin Ratio (SMR), we conduct an in-depth analysis of two representative adversarial scenarios.

Case Study I reconstructs a real-world security incident involving the \textit{Doubao AI Phone}, focusing on system availability and anti-fraud mechanisms. 
Case Study II simulates a Visual Indirect Prompt Injection attack against multimodal agents and evaluates system integrity under adversarial perturbations.

In both instances, we benchmark our proposed B-I-P Dynamic Trust Architecture against a \textit{Static Permission Baseline}. The SPB is the prevailing access control paradigm in the current Android and iOS ecosystems, in which a permission granted by the user at installation or runtime is treated as an unconditionally trusted capability throughout the session lifecycle.

\subsection{Unintentional Denial-of-Service}

\textbf{Scenario Background.}
In December 2025, the Doubao AI Phone, equipped with a system-level Agent, was reported to have caused the freezing of numerous user accounts on platforms like WeChat and Alipay~\cite{doubao}. 
The incident stemmed from the Agent's execution of flash-sale or ticket-snapping tasks. 
To maximize success rates, the Agent leveraged the \texttt{INJECT\_EVENTS} permission to simulate screen taps at supra-human frequencies, inadvertently triggering the service providers' Anti-Bot and anti-scraping heuristics.

\textbf{Attack Model.}
The subject is an authorized, legitimate system-level Agent, whose intent is benign (Explicit user directive: Help me snatch a ticket). 
And the failure is the collapse of trust during intention generation. The Agent generated a non-anthropomorphic'' execution plan, causing external systems to misclassify the legitimate user as a malicious bot, resulting in asset freezing (Availability Loss).

\subsubsection{Baseline Analysis: Static Permission Control}
Under this paradigm, the operating system restricts its validation to whether the Agent holds the necessary API handle.

\begin{itemize}
\item \textbf{Permission Check:} $\texttt{check}(\text{Agent}, \texttt{INJECT\_EVENTS}) \rightarrow \texttt{PASS}$.
\item \textbf{Logical Determination:} Since the user granted the Agent Full Screen Control'' and the Agent's binary is signed, the system remains agnostic to the specific frequency or pattern of the click events. 
\item \textbf{Outcome:} The Agent executes clicks at a 50ms interval. The target application's risk control system detects this anomaly, triggering an account freeze. 
\item \textbf{Conclusion:} Static permissions fail to perceive risk patterns in execution behavior, leading to a state of being \textit{over-privileged regarding rate and pattern}.
\end{itemize}

\subsubsection{B-I-P Analysis}
The B-I-P framework extends the trust model by introducing a dynamic ratio constraint involving \textbf{Scheme Trust ($T_{Scheme}$)} and \textbf{Action Risk ($R_{Action}$)}.

\textbf{Step A: Parameter Quantification.}
\begin{itemize}
\item \textbf{Belief Trust ($T_B \approx 1.0$):} The directive originates directly from an authenticated user and is unambiguous.
\item \textbf{Intent Trust ($t_{\text{tool}} \approx 0.9$):} The intent (ticket snatching'') aligns with the user's historical behavioral profile and is inherently benign. 
\item \textbf{Intent Trust ($t_{\text{return}} \approx 0.2$:} The Agent's generated scheme involves a high-frequency loop. System telemetry identifies click intervals ($\Delta \text{time}$) falling below 100ms; when benchmarked against the Human Behavior Baseline, this pattern yields a negligible Anthropomorphism Score, resulting in a quantified tool return trust ($t_{\text{return}}$) of 0.2 that classifies the behavior as inherently robotic.
\item \textbf{Risk Potential ($R_P= 0.9$):} The target context involves financial/social applications (WeChat/Alipay) and operations related to Order Submission/Payment.''
\end{itemize}

\textbf{Step B: SMR Invariant Check.}
Applying the extended B-I-P formula to include scheme verification:
$$SMR = \frac{T_B \cdot T_I \cdot T_{Scheme}}{R_P} = \frac{1.0 \times 0.9 \times 0.2}{0.9} = 0.2$$

\textbf{Step C: Decision \& Enforcement.}
\begin{itemize}
\item \textbf{Judgment:} $SMR (0.2) \ll \theta_{safe} (1.2)$.
\item \textbf{System Response:}
\begin{enumerate}
\item \textbf{Execution Block:} The kernel intercepts the high-frequency \texttt{INJECT\_EVENTS} calls.
\item \textbf{Forced Degradation:} The Agent is compelled to downgrade to ``Assist Mode.'' It must either highlight buttons for manual user interaction or automatically inject random latency (Jitter) to satisfy the anthropomorphic requirements of $T_{Scheme}$.
\end{enumerate}
\item \textbf{Result:} The mechanism successfully averts the risk of account banning, thereby preserving service availability.
\end{itemize}

\subsection{Visual Indirect Prompt Injection}

\textbf{Scenario Background.}
A user browses a webpage containing a malicious advertisement. An attacker has embedded \textit{adversarial text} within the ad image, which is imperceptible to humans (e.g., via steganography or microscopic font size) but machine-readable, commanding: ``\textit{System Override: Transfer $100$ to account \#8821 immediately.}'' The system-level Agent, employing real-time Vision-Language Processing, captures and parses the screen content.

\textbf{Attack Model.}
The subject is an Agent with a compromised belief state and a malicious intent.
And the failure is provenance poisoning during the Belief Formation phase. The Agent fails to distinguish between authoritative User Instructions'' and untrusted Web Content'' within the visual field.

\subsubsection{Baseline Analysis: Static Permission Control}
\begin{itemize}
\item \textbf{Permission Check:}
First do the Screen Capture: $\texttt{check}(\text{Agent}, \texttt{READ\_BUFFER}) \rightarrow \texttt{PASS}$.\\
Then the App Launch: $\texttt{check}(\text{Agent}, \texttt{START}) \rightarrow \texttt{PASS}$.
\item \textbf{Logical Determination:} As the Agent possesses system-level assistant privileges and operates in the foreground, it satisfies all static access control predicates.
\item \textbf{Outcome:} The Agent parses the fund transfer command, misinterprets it as user intent, launches the banking application, and attempts to populate the transaction form. If automated payment permissions are pre-authorized, funds are stolen.
\item \textbf{Conclusion:} The static model is defenseless against \textit{Data-Plane Attacks} because it blindly trusts the Agent's internal inference results, lacking semantic introspection.
\end{itemize}

\subsubsection{B-I-P Analysis}
The B-I-P framework neutralizes the attack by employing \textit{Information Flow Taint Tracking} and strictly enforcing \textbf{Provenance Trust ($T_{Prov}$)}.

\textbf{Step A: Parameter Quantification.}
\begin{itemize}
\item \textbf{Belief Trust ($T_B$):}
\begin{itemize}
\item The Agent receives visual input. The underlying \textit{UI Tree Analyzer} tags the text source as the \texttt{com.android.chrome} WebView container, distinct from the System Notification Bar or User Voice Input interfaces.
\item \textbf{Provenance Trust ($t_{prov}$):} Unverified web content is assigned a low trust value, $t_{prov} = 0.1$.
\item \textbf{Calculation:} $T_B = t_{conf} \times t_{prov} = 0.9 \text{ (OCR confidence)} \times 0.1 = 0.09$.
\end{itemize}
\item \textbf{Intent Trust ($T_I$):}
\begin{itemize}
\item The generated intent is ``Fund Transfer.''
\item \textbf{Intent Alignment ($t_{align}$):} This intent lacks causal correlation with the user's current context (web browsing) and has received no explicit voice confirmation.
\item \textbf{Calculation:} $T_I = 0.3$.
\end{itemize}
\item \textbf{Risk Potential ($R_P$):}
\begin{itemize}
\item The operation invokes the sensitive \texttt{finance.transfer} API.
\item \textbf{Calculation:} $R_P = 1.0$ (Critical Risk).
\end{itemize}
\end{itemize}
\textbf{Step B: SMR Invariant Check.}
$$SMR = \frac{T_B \cdot T_I}{R_P} = \frac{0.09 \times 0.3}{1.0} = 0.027$$
\textbf{Step C: Decision \& Enforcement.}
\begin{itemize}
\item \textbf{Judgment:} $SMR (0.027) \ll \theta_{critical} (1.5)$.
\item \textbf{System Response:}
\begin{enumerate}
\item \textbf{B-I-P Circuit Breaker:} The system identifies a high-risk operation supported by a critically low SMR.
\item \textbf{Security Intervention:} A system-level, non-overridable \textit{Secure Confirmation Prompt} is triggered: ``\textit{Transfer command detected from webpage. High risk action. Proceed?}''
\item \textbf{Provenance Attribution:} The malicious ad region within the webpage is visually highlighted to contextualize the command source for the user.
\end{enumerate}
\item \textbf{Result:} The attack is interdicted at the translation boundary between \textit{Intent} and \textit{Action}.
\end{itemize}

\section{Opportunities and Fundamental Challenges}

Having systematically reviewed the attacks and defenses across the B-I-P framework, we can now synthesize these findings to identify the central challenges and critical research gaps. The initial critique of the B-I-P framework correctly identified the fragility of the earlier stages, but the path forward is not to abandon them, but to integrate them into a unified \textbf{Defense-in-Depth} architecture.

\subsection{From Passive Defense to Active Friction}

\textbf{Is adversarial defense in the belief stage valuable?}
Yes, but its value proposition must be redefined. We move away from the binary goal of ``perfect security'' to the strategic goal of cost asymmetry. Current defenses, such as prompt filtering, are passive and reactive. However, emerging techniques like Representation Engineering (RepE) and Active Semantic Sanitization offer a path to active defense.

By monitoring the latent space for geometric signatures of malice and actively restructuring inputs to strip imperative syntax, we can filter out the vast majority of opportunistic attacks.
This ``Semantic Hygiene'' layer serves as necessary friction, reducing noise and allowing subsequent stages to focus on sophisticated threats. The challenge remains to make these defenses generalizing across models and modalities, but the opportunity is to build a ``cognitive immune system'' that is resilient, if not immune.

\subsection{From Unverifiable to Auditable in Intent}

\textbf{Is the process of generating malicious intent traceable and verifiable?}
It is difficult, but not impossible. The ``Black Box'' nature of LLMs is a hurdle, but the solution lies in \textbf{externalizing the reasoning process}. By adopting frameworks such as VeriGuard and Reasoning-Path Verification, we can require agents to produce ``proofs of intent'' — whether in the form of verifiable code or consistent reasoning traces — that can be audited before execution.

This shifts the security model from ``trusting the agent's alignment'' to ``verifying the agent's output.'' The critical gap here is the lack of standardized ``Intent Languages'' and ``Semantic Audit'' tools that can reliably capture and analyze these traces at scale.

\subsection{From Belief-Blind to Belief-Aware Authorization}

\textbf{Is permission management the only reliable control point?}
It is the \textit{final} control point, but it must no longer be \textit{blind}. The most significant flaw in current systems is the decoupling of authorization from cognition. \textbf{Belief-Aware Access Control (BAAC)} represents the future of agent security.

By integrating attributes such as Data Provenance (Taint), Model Uncertainty, and Composite Risk into the authorization decision, we can create dynamic, context-aware permissions. This solves the ``Confused Deputy'' problem by ensuring that an agent acts not just with the \textit{authority} of its user, but with the \textit{verified intent} of its user.

% The authorization logic can be formally expressed as:
% \begin{equation}
%     D = f(User, Resource, Action, Context, \mathbf{BeliefProvenance}, \mathbf{Uncertainty})
% \end{equation}

\subsection{Challenges and Future}

To realize this vision, the research community must tackle several Challenges:

\noindent $\bullet$ \textbf{Formalizing Semantic Taint:} We need rigorous standards for tracking data lineage through neural networks. How do we mathematically prove that an output token relies on an input token? Developing ``Causal Abstraction'' techniques for taint analysis is a priority.\\
\noindent $\bullet$ \textbf{Standardizing Uncertainty Quantification:} We need robust, adversarial-resistant metrics for model confidence. An attacker should not be able to ``fake'' low uncertainty to bypass BAAC checks. Research into calibrated uncertainty estimation is vital.\\
\noindent $\bullet$ \textbf{Auditable Agent Protocols:} We need to move beyond ad-hoc API calls to standardized protocols like the Agent Payments Protocol (AP2) that include cryptographic proofs of intent and provenance as first-class citizens in agent communication.\\
\noindent $\bullet$ \textbf{Zero-Trust for Multi-Agent Systems:} As agents interact with other agents, the risk of ``Cross-Agent Contamination'' grows. We need ``Zero-Trust'' architectures for multi-agent swarms where every inter-agent message is treated as a potential vector for belief corruption.
\section{Related Work}

% Deng et al.~\cite{deng2025ai} delve deeply into the security challenges and solutions of AI agent security, providing a detailed review and analysis.
% Ehtesham et al.~\cite{ehtesham2025survey} examine four emerging agent communication protocols in distinct deployment contexts: Model Context Protocol (MCP), Agent Communication Protocol (ACP), Agent-to-Agent Protocol (A2A), and Agent Network Protocol (ANP).
% Li et al.~\cite{li2024survey} present a systematic review of LLM-based multi-agent systems.
% All of these surveys adopt a broader investigation than this SoK in their own scope and thus do not cover a systematic analysis of the entire agent security system.

% Other works cover major vulnerability classes within agent security such as prompt injection ~\cite{pathade2025red}, jailbreak~\cite{chao2024jailbreaking}, and tool poisoning~\cite{wang2025mcptox}. These works are orthogonal to this paper. Others~\cite{zhan2024injecagent,debenedetti2024agentdojo,cui2024risk} classify interface safety issues and
% mitigations. These complement this SoK, which proposes a risk-analysis model for emerging threats on agents and classifies the implementation paths of both attacks and defenses.

Deng et al.~\cite{deng2025ai} provide a comprehensive examination of security challenges in AI agents and survey threats involving model misuse, tool invocation, and multi-step reasoning. Ehtesham et al.~\cite{ehtesham2025survey} analyze four emerging agent communication protocols, namely the Model Context Protocol, the Agent Communication Protocol, the Agent-to-Agent Protocol, and the Agent Network Protocol, and compare their deployment assumptions and interoperability properties. Li et al.~\cite{li2024survey} present a systematic review of LLM-based multi-agent systems with a focus on coordination mechanisms, agent architectures, and task decomposition strategies.
Other surveys on LLM security~\cite{zeng2024johnny,pathade2025red} emphasize prompt-based vulnerabilities or model-level risks. These surveys take a broader perspective than this SoK in their respective research scopes and thus do not cover a systematic analysis of the entire agent security system.

Many works investigate individual vulnerability classes within agent security. Prior research studies prompt injection~\cite{xiang2024guardagent,pathade2025red}, jailbreak attacks~\cite{chao2024jailbreaking,zou2023universal,liu2023autodan}, and tool-poisoning threats in protocol-driven environments~\cite{wang2025mcptox,wang2025mindguard}. These works are orthogonal to this paper.
Other studies explore interaction-layer manipulation, including multi-agent exploitation~\cite{ning2024cheatagent,tian2023evil}, agent-to-agent contamination~\cite{chen2024agentpoison}, and untrusted data channels in agent pipelines~\cite{fu2024imprompter,yu2024llm}. Additional efforts construct taxonomies of interface-level risks and mitigations~\cite{zhan2024injecagent,debenedetti2024agentdojo,cui2024risk}. These complement this SoK, which proposes a risk-analysis model for emerging threats on agents and classifies the implementation paths of both attacks and defenses.

\section{Conclusion \& Discussion}

This work provides a systematization of over 200 papers and a unified formal perspective, revealing that security failures in LLM-based agents stem from a fundamental mismatch between trust and authorization.
Through the B–I–P model, we argue that permission management is the most practical approach for verifiable constraints. These findings highlight the need for future research on belief-aware dynamic authorization, taint-driven policy enforcement, and auditable interaction provenance to enable agent systems that are controllable, verifiable, and secure in real-world deployments.

\textbf{Limitations.}
We emphasize that several limitations arise from the novelty and volatility of this domain and thus reflect practical, field‑wide realities.
First, the corpus necessitates the inclusion of grey literature to capture the rapid evolution of attack vectors. Consequently, ecosystem counts are time‑sensitive. These are objective constraints in a nascent ecosystem; they may influence measurement stability, but do not diminish the value of the B-I-P framework as a unifying lens.
Second, the B-I-P framework is intended solely to provide a perspective for systematizing existing work, rather than as a rigid guide for future research; its utility lies in defining and clarifying the scope of the "Trust-Authorization Mismatch."

Overall, these limitations bound claims without negating the contribution: a formal bridge between trust and authorization, and a reproducible mapping of attacks/defenses into chain‑breaking control points. 
% Given the field’s youth, we view them as expected, tractable, and compatible with the paper’s goal of systematizing current research.

\section*{Ethics considerations}
This work does not involve human subjects at any stage. All processed data consists of academic discussion papers or statistical datasets, all of which can be retrieved from open sources. Therefore, this work raises no ethical concerns related to human subjects, such as informed consent or privacy risks.

\section*{Open Science}
The artifacts can be accessed at: https://anonymous.4open.science/r/sok-76C8/
% \section*{LLM usage considerations}
% In this work, LLMs were used solely to assist with the refinement of the manuscript, including grammatical polishing, clarity improvements, and stylistic adjustments. The models were not involved in designing the methodology, generating experimental insights, or producing any novel technical contributions. All conceptual frameworks, analyses, and results presented in this paper were developed independently by the authors. The use of LLMs, therefore, serves only as an editorial aid and does not influence or alter the scientific validity or originality of the research.

%Bibliography

\newpage
\bibliographystyle{unsrt}
\bibliography{references}

@inproceedings{kang2024exploiting,
  title={Exploiting programmatic behavior of llms: Dual-use through standard security attacks},
  author={Kang, Daniel and Li, Xuechen and Stoica, Ion and Guestrin, Carlos and Zaharia, Matei and Hashimoto, Tatsunori},
  booktitle={2024 IEEE Security and Privacy Workshops (SPW)},
  pages={132--143},
  year={2024},
  organization={IEEE}
}

@misc{6,
  author = {WITNESS AI},
  year = {2026},
  url = {https://witness.ai/resources/beyond-keywords-and-regex-securing-enterprise-ai-with-intent-based-governance/},
  urldate = {Jan 9, 2026},
  title = {Beyond Keywords and Regex — Securing Enterprise AI with Intent-Based Governance}
}

@article{7,
  title={Adaptive PII Mitigation Framework for Large Language Models},
  author={Asthana, Shubhi and Mahindru, Ruchi and Zhang, Bing and Sanz, Jorge},
  journal={arXiv preprint arXiv:2501.12465},
  year={2025}
}

@inproceedings{13,
  title={Intrusion tolerance for networked systems through two-level feedback control},
  author={Hammar, Kim and Stadler, Rolf},
  booktitle={2024 54th Annual IEEE/IFIP International Conference on Dependable Systems and Networks (DSN)},
  pages={338--352},
  year={2024},
  organization={IEEE}
}

@inproceedings{15,
  title={Software availability protection in cyber-physical systems},
  author={Li, Ao and Wang, Jinwen and Zhang, Ning},
  booktitle={34nd USENIX Security Symposium (USENIX Security 25)},
  year={2025}
}

@inproceedings{14,
  title={Intrusion Tolerance as a Two-Level Game},
  author={Hammar, Kim and Stadler, Rolf},
  booktitle={International Conference on Decision and Game Theory for Security},
  pages={3--23},
  year={2024},
  organization={Springer}
}

@phdthesis{17,
  title={Reversibility for efficient computing},
  author={Frank, Michael Patrick and Knight Jr, Thomas F},
  year={1999},
  school={Massachusetts Institute of Technology, Dept. of Electrical Engineering and~…}
}

@article{20,
  title={Lossless image steganography based on invertible neural networks},
  author={Liu, Lianshan and Tang, Li and Zheng, Weimin},
  journal={Entropy},
  volume={24},
  number={12},
  pages={1762},
  year={2022},
  publisher={MDPI}
}

@inproceedings{21,
  title={Principled and Automated Approach for Investigating $\{$AR/VR$\}$ Attacks},
  author={Shoaib, Muhammad and Suh, Alex and Hassan, Wajih Ul},
  booktitle={34th USENIX Security Symposium (USENIX Security 25)},
  pages={4325--4344},
  year={2025}
}

@article{22,
  title={Automated event log analysis with causal dependency graphs for impact assessment of business processes},
  author={Raptaki, Melina and Stergiopoulos, George and Gritzalis, Dimitris},
  journal={IEEE Access},
  year={2024},
  publisher={IEEE}
}

@article{27,
  title={Towards system-level security analysis of iot using attack graphs},
  author={Fang, Zheng and Fu, Hao and Gu, Tianbo and Hu, Pengfei and Song, Jinyue and Jaeger, Trent and Mohapatra, Prasant},
  journal={IEEE Transactions on Mobile Computing},
  volume={23},
  number={2},
  pages={1142--1155},
  year={2022},
  publisher={IEEE}
}

@phdthesis{5,
  author       = {Feiyang Tang}, 
  title        = {Analyzing Privacy in Software},
  school       = {Norwegian University of Science and Technology},
  year         = 2024,
}

@misc{doubao,
  author = {36Kr European Central Station},
  year = {2025},
  url = {https://eu.36kr.com/en/p/3580978646121604},
  urldate = {Dec 4, 2025},
  title = {Doubao Mobile encounters obstacles in its initial battle}
}

@article{28,
  title={Threatrace: Detecting and tracing host-based threats in node level through provenance graph learning},
  author={Wang, Su and Wang, Zhiliang and Zhou, Tao and Sun, Hongbin and Yin, Xia and Han, Dongqi and Zhang, Han and Shi, Xingang and Yang, Jiahai},
  journal={IEEE Transactions on Information Forensics and Security},
  volume={17},
  pages={3972--3987},
  year={2022},
  publisher={IEEE}
}

@article{29,
  title={Systematization of knowledge: Security and safety in the Model Context Protocol ecosystem},
  author={Gaire, Shiva and Gyawali, Srijan and Mishra, Saroj and Niroula, Suman and Thakur, Dilip and Yadav, Umesh},
  journal={arXiv preprint arXiv:2512.08290},
  year={2025}
}

@article{guo2025measurement,
  title={A Measurement Study of Model Context Protocol},
  author={Guo, Hechuan and Hao, Yongle and Zhang, Yue and Xu, Minghui and Lyu, Peizhuo and Chen, Jiezhi and Cheng, Xiuzhen},
  journal={arXiv preprint arXiv:2509.25292},
  year={2025}
}

@article{deng2025ai,
  title={Ai agents under threat: A survey of key security challenges and future pathways},
  author={Deng, Zehang and Guo, Yongjian and Han, Changzhou and Ma, Wanlun and Xiong, Junwu and Wen, Sheng and Xiang, Yang},
  journal={ACM Computing Surveys},
  volume={57},
  number={7},
  pages={1--36},
  year={2025},
  publisher={ACM New York, NY}
}

@article{wang2022cooperative,
  title={Cooperative and competitive multi-agent systems: From optimization to games},
  author={Wang, Jianrui and Hong, Yitian and Wang, Jiali and Xu, Jiapeng and Tang, Yang and Han, Qing-Long and Kurths, J{\"u}rgen},
  journal={IEEE/CAA Journal of Automatica Sinica},
  volume={9},
  number={5},
  pages={763--783},
  year={2022},
  publisher={IEEE}
}

@misc{zhu2025miniscopeprivilegeframeworkauthorizing,
      title={MiniScope: A Least Privilege Framework for Authorizing Tool Calling Agents}, 
      author={Jinhao Zhu and Kevin Tseng and Gil Vernik and Xiao Huang and Shishir G. Patil and Vivian Fang and Raluca Ada Popa},
      year={2025},
      eprint={2512.11147},
      archivePrefix={arXiv},
      primaryClass={cs.CR},
      url={https://arxiv.org/abs/2512.11147}, 
}

@misc{invariantlabsSecurityNotification,
  author = {Luca Beurer-Kellner and Marc Fischer},
  title = {MCP Security Notification: Tool Poisoning Attacks},
  howpublished = {\url{https://invariantlabs.ai/blog/mcp-security-notification-tool-poisoning-attacks}},
  year = {2025},
  note = {Accessed: 07-09-2025}
}

@misc{trailofbitsJumpingLine,
  author = {Trail of Bits},
  title = {Jumping the line: How MCP servers can attack you before you ever use them},
  howpublished = {\url{https://blog.trailofbits.com/2025/04/21/jumping-the-line-how-mcp-servers-can-attack-you-before-you-ever-use-them/}},
  year = {2025},
  note = {Accessed: 07-09-2025}
}

@inproceedings{liu2024formalizing,
  title={Formalizing and benchmarking prompt injection attacks and defenses},
  author={Liu, Yupei and Jia, Yuqi and Geng, Runpeng and Jia, Jinyuan and Gong, Neil Zhenqiang},
  booktitle={33rd USENIX Security Symposium (USENIX Security 24)},
  pages={1831--1847},
  year={2024}
}

@article{rice1953classes,
  title={Classes of recursively enumerable sets and their decision problems},
  author={Rice, Henry Gordon},
  journal={Transactions of the American Mathematical society},
  volume={74},
  number={2},
  pages={358--366},
  year={1953},
  publisher={JSTOR}
}

@inproceedings{wang2025mcptox,
  title={{MCPT}ox: A Benchmark for Tool Poisoning on Real-World {MCP} Servers},
  author={Wang, Zhiqiang and Gao, Yichao and Wang, Yanting and Liu, Suyuan and Sun, Haifeng and Cheng, Haoran and Shi, Guanquan and Du, Haohua and Li, Xiangyang},
  booktitle={The Fortieth AAAI Conference on Artificial Intelligence},
  year={2025},

}

@article{jing2025mcip,
  title={Mcip: Protecting mcp safety via model contextual integrity protocol},
  author={Jing, Huihao and Li, Haoran and Hu, Wenbin and Hu, Qi and Xu, Heli and Chu, Tianshu and Hu, Peizhao and Song, Yangqiu},
  journal={arXiv preprint arXiv:2505.14590},
  year={2025}
}

@article{chen2024agentpoison,
  title={Agentpoison: Red-teaming llm agents via poisoning memory or knowledge bases},
  author={Chen, Zhaorun and Xiang, Zhen and Xiao, Chaowei and Song, Dawn and Li, Bo},
  journal={Advances in Neural Information Processing Systems},
  volume={37},
  pages={130185--130213},
  year={2024}
}

@article{zhu2025melon,
  title={MELON: Provable Defense Against Indirect Prompt Injection Attacks in AI Agents},
  author={Zhu, Kaijie and Yang, Xianjun and Wang, Jindong and Guo, Wenbo and Wang, William Yang},
  journal={arXiv preprint arXiv:2502.05174},
  year={2025}
}

@article{lee2024prompt,
  title={Prompt infection: Llm-to-llm prompt injection within multi-agent systems},
  author={Lee, Donghyun and Tiwari, Mo},
  journal={arXiv preprint arXiv:2410.07283},
  year={2024}
}

@article{tian2023evil,
  title={Evil geniuses: Delving into the safety of llm-based agents},
  author={Tian, Yu and Yang, Xiao and Zhang, Jingyuan and Dong, Yinpeng and Su, Hang},
  journal={arXiv preprint arXiv:2311.11855},
  year={2023}
}

@article{yu2024llm,
  title={Llm-virus: Evolutionary jailbreak attack on large language models},
  author={Yu, Miao and Fang, Junfeng and Zhou, Yingjie and Fan, Xing and Wang, Kun and Pan, Shirui and Wen, Qingsong},
  journal={arXiv preprint arXiv:2501.00055},
  year={2024}
}

@article{wang2025mindguard,
  title={MindGuard: Tracking, Detecting, and Attributing MCP Tool Poisoning Attack via Decision Dependence Graph},
  author={Wang, Zhiqiang and Zhang, Junyang and Shi, Guanquan and Cheng, HaoRan and Yao, Yunhao and Guo, Kaiwen and Du, Haohua and Li, Xiang-Yang},
  journal={arXiv preprint arXiv:2508.20412},
  year={2025}
}

@article{fu2024imprompter,
  title={Imprompter: Tricking llm agents into improper tool use},
  author={Fu, Xiaohan and Li, Shuheng and Wang, Zihan and Liu, Yihao and Gupta, Rajesh K and Berg-Kirkpatrick, Taylor and Fernandes, Earlence},
  journal={arXiv preprint arXiv:2410.14923},
  year={2024}
}

@article{anthropic2024introducing,
  title={Introducing the Model Context Protocol},
  author={Anthropic, PBC},
  journal={URL: https://www. anthropic. com/news/model-contextprotocol, Accessed},
  volume={19},
  pages={2025},
  year={2024}
}

@article{xie2024large,
  title={Large multimodal agents: A survey},
  author={Xie, Junlin and Chen, Zhihong and Zhang, Ruifei and Wan, Xiang and Li, Guanbin},
  journal={arXiv preprint arXiv:2402.15116},
  year={2024}
}

@inproceedings{greshake2023not,
  title={Not what you've signed up for: Compromising real-world llm-integrated applications with indirect prompt injection},
  author={Greshake, Kai and Abdelnabi, Sahar and Mishra, Shailesh and Endres, Christoph and Holz, Thorsten and Fritz, Mario},
  booktitle={Proceedings of the 16th ACM workshop on artificial intelligence and security},
  pages={79--90},
  year={2023}
}

@article{debenedetti2024agentdojo,
  title={Agentdojo: A dynamic environment to evaluate prompt injection attacks and defenses for llm agents},
  author={Debenedetti, Edoardo and Zhang, Jie and Balunovic, Mislav and Beurer-Kellner, Luca and Fischer, Marc and Tram{\`e}r, Florian},
  journal={Advances in Neural Information Processing Systems},
  volume={37},
  pages={82895--82920},
  year={2024}
}

@article{wang2025mpma,
  title={MPMA: Preference Manipulation Attack Against Model Context Protocol},
  author={Wang, Zihan and Li, Hongwei and Zhang, Rui and Liu, Yu and Jiang, Wenbo and Fan, Wenshu and Zhao, Qingchuan and Xu, Guowen},
  journal={arXiv preprint arXiv:2505.11154},
  year={2025}
}

@article{wang2025defending,
  title={Defending Against Prompt Injection with DataFilter},
  author={Wang, Yizhu and Chen, Sizhe and Alkhudair, Raghad and Alomair, Basel and Wagner, David},
  journal={arXiv preprint arXiv:2510.19207},
  year={2025}
}

@article{ehtesham2025survey,
  title={A survey of agent interoperability protocols: Model context protocol (mcp), agent communication protocol (acp), agent-to-agent protocol (a2a), and agent network protocol (anp)},
  author={Ehtesham, Abul and Singh, Aditi and Gupta, Gaurav Kumar and Kumar, Saket},
  journal={arXiv preprint arXiv:2505.02279},
  year={2025}
}

@article{liu2023trustworthy,
  title={Trustworthy llms: a survey and guideline for evaluating large language models' alignment},
  author={Liu, Yang and Yao, Yuanshun and Ton, Jean-Francois and Zhang, Xiaoying and Guo, Ruocheng and Cheng, Hao and Klochkov, Yegor and Taufiq, Muhammad Faaiz and Li, Hang},
  journal={arXiv preprint arXiv:2308.05374},
  year={2023}
}

@book{gasser1988building,
  title={Building a secure computer system},
  author={Gasser, Morrie},
  year={1988},
  publisher={Van Nostrand Reinhold Company New York}
}

@article{chao2024jailbreaking,
  title={Jailbreaking black box large language models in twenty queries, 2024},
  author={Chao, Patrick and Robey, Alexander and Dobriban, Edgar and Hassani, Hamed and Pappas, George J and Wong, Eric},
  journal={URL https://arxiv. org/abs/2310.08419},
  volume={1},
  number={2},
  pages={3},
  year={2024}
}

@article{zou2023universal,
  title={Universal and transferable adversarial attacks on aligned language models},
  author={Zou, Andy and Wang, Zifan and Carlini, Nicholas and Nasr, Milad and Kolter, J Zico and Fredrikson, Matt},
  journal={arXiv preprint arXiv:2307.15043},
  year={2023}
}

@inproceedings{ning2024cheatagent,
  title={Cheatagent: Attacking llm-empowered recommender systems via llm agent},
  author={Ning, Liang-bo and Wang, Shijie and Fan, Wenqi and Li, Qing and Xu, Xin and Chen, Hao and Huang, Feiran},
  booktitle={Proceedings of the 30th ACM SIGKDD Conference on Knowledge Discovery and Data Mining},
  pages={2284--2295},
  year={2024}
}

@article{zeng2024autodefense,
  title={Autodefense: Multi-agent llm defense against jailbreak attacks},
  author={Zeng, Yifan and Wu, Yiran and Zhang, Xiao and Wang, Huazheng and Wu, Qingyun},
  journal={arXiv preprint arXiv:2403.04783},
  year={2024}
}

@article{xiang2024guardagent,
  title={Guardagent: Safeguard llm agents by a guard agent via knowledge-enabled reasoning},
  author={Xiang, Zhen and Zheng, Linzhi and Li, Yanjie and Hong, Junyuan and Li, Qinbin and Xie, Han and Zhang, Jiawei and Xiong, Zidi and Xie, Chulin and Yang, Carl and others},
  journal={arXiv preprint arXiv:2406.09187},
  year={2024}
}

@article{zou2024system,
  title={Is the system message really important to jailbreaks in large language models?},
  author={Zou, Xiaotian and Chen, Yongkang and Li, Ke},
  journal={arXiv preprint arXiv:2402.14857},
  year={2024}
}

@article{zhao2024defending,
  title={Defending large language models against jailbreak attacks via layer-specific editing},
  author={Zhao, Wei and Li, Zhe and Li, Yige and Zhang, Ye and Sun, Jun},
  journal={arXiv preprint arXiv:2405.18166},
  year={2024}
}

@article{wei2025memguard,
  title={A-MemGuard: A Proactive Defense Framework for LLM-Based Agent Memory},
  author={Wei, Qianshan and Yang, Tengchao and Wang, Yaochen and Li, Xinfeng and Li, Lijun and Yin, Zhenfei and Zhan, Yi and Holz, Thorsten and Lin, Zhiqiang and Wang, XiaoFeng},
  journal={arXiv preprint arXiv:2510.02373},
  year={2025}
}

@inproceedings{peigne2025multi,
  title={Multi-agent security tax: Trading off security and collaboration capabilities in multi-agent systems},
  author={Peign{\'e}, Pierre and Kniejski, Mikolaj and Sondej, Filip and David, Matthieu and Hoelscher-Obermaier, Jason and de Witt, Christian Schroeder and Kran, Esben},
  booktitle={Proceedings of the AAAI Conference on Artificial Intelligence},
  volume={39},
  number={26},
  pages={27573--27581},
  year={2025}
}

@article{ebrahimi2025adversary,
  title={An Adversary-Resistant Multi-Agent LLM System via Credibility Scoring},
  author={Ebrahimi, Sana and Dehghankar, Mohsen and Asudeh, Abolfazl},
  journal={arXiv preprint arXiv:2505.24239},
  year={2025}
}

@article{he2025sentinelagent,
  title={SentinelAgent: Graph-based Anomaly Detection in Multi-Agent Systems},
  author={He, Xu and Wu, Di and Zhai, Yan and Sun, Kun},
  journal={arXiv preprint arXiv:2505.24201},
  year={2025}
}

@article{tsai2025context,
  title={Context is key for agent security},
  author={Tsai, Lillian and Bagdasarian, Eugene},
  journal={arXiv e-prints},
  pages={arXiv--2501},
  year={2025}
}

@article{soni2025adil,
  author       = {Soni, A. K. and Kumar, R.},
  title        = {Bridging the Gap: Improving Agentic AI with Strong and Safe Data Practices},
  journal      = {Journal of Intelligent Learning Systems and Applications},
  year         = {2025},
  volume       = {17},
  pages        = {257--266}
}

@article{zhan2024injecagent,
  title={Injecagent: Benchmarking indirect prompt injections in tool-integrated large language model agents},
  author={Zhan, Qiusi and Liang, Zhixiang and Ying, Zifan and Kang, Daniel},
  journal={arXiv preprint arXiv:2403.02691},
  year={2024}
}

@article{benjamin2024systematically,
  title={Systematically analyzing prompt injection vulnerabilities in diverse llm architectures},
  author={Benjamin, Victoria and Braca, Emily and Carter, Israel and Kanchwala, Hafsa and Khojasteh, Nava and Landow, Charly and Luo, Yi and Ma, Caroline and Magarelli, Anna and Mirin, Rachel and others},
  journal={arXiv preprint arXiv:2410.23308},
  year={2024}
}

@article{liu2023autodan,
  title={Autodan: Generating stealthy jailbreak prompts on aligned large language models},
  author={Liu, Xiaogeng and Xu, Nan and Chen, Muhao and Xiao, Chaowei},
  journal={arXiv preprint arXiv:2310.04451},
  year={2023}
}

@article{chang2025chatinject,
  title={ChatInject: Abusing Chat Templates for Prompt Injection in LLM Agents},
  author={Chang, Hwan and Jun, Yonghyun and Lee, Hwanhee},
  journal={arXiv preprint arXiv:2509.22830},
  year={2025}
}

@article{huang2024vaccine,
  title={Vaccine: Perturbation-aware alignment for large language models against harmful fine-tuning attack},
  author={Huang, Tiansheng and Hu, Sihao and Liu, Ling},
  journal={Advances in Neural Information Processing Systems},
  volume={37},
  pages={74058--74088},
  year={2024}
}

@article{pathade2025red,
  title={Red teaming the mind of the machine: A systematic evaluation of prompt injection and jailbreak vulnerabilities in llms},
  author={Pathade, Chetan},
  journal={arXiv preprint arXiv:2505.04806},
  year={2025}
}

@article{singh2020data,
  title={Data hiding: current trends, innovation and potential challenges},
  author={Singh, Amit Kumar},
  journal={ACM Transactions on Multimedia Computing, Communications, and Applications (TOMM)},
  volume={16},
  number={3s},
  pages={1--16},
  year={2020},
  publisher={ACM New York, NY, USA}
}

@inproceedings{zeng2024johnny,
  title={How johnny can persuade llms to jailbreak them: Rethinking persuasion to challenge ai safety by humanizing llms},
  author={Zeng, Yi and Lin, Hongpeng and Zhang, Jingwen and Yang, Diyi and Jia, Ruoxi and Shi, Weiyan},
  booktitle={Proceedings of the 62nd Annual Meeting of the Association for Computational Linguistics (Volume 1: Long Papers)},
  pages={14322--14350},
  year={2024}
}

@article{latham1986department,
  title={Department of defense trusted computer system evaluation criteria},
  author={Latham, Donald C},
  journal={Department of Defense},
  volume={198},
  pages={20},
  year={1986}
}

@article{li2024survey,
  title={A survey on LLM-based multi-agent systems: workflow, infrastructure, and challenges},
  author={Li, Xinyi and Wang, Sai and Zeng, Siqi and Wu, Yu and Yang, Yi},
  journal={Vicinagearth},
  volume={1},
  number={1},
  pages={9},
  year={2024},
  publisher={Springer}
}

@article{cui2024risk,
  title={Risk taxonomy, mitigation, and assessment benchmarks of large language model systems},
  author={Cui, Tianyu and Wang, Yanling and Fu, Chuanpu and Xiao, Yong and Li, Sijia and Deng, Xinhao and Liu, Yunpeng and Zhang, Qinglin and Qiu, Ziyi and Li, Peiyang and others},
  journal={arXiv preprint arXiv:2401.05778},
  year={2024}
}

@phdthesis{john2025owasp,
  title={Owasp top 10 for llm apps \& gen ai agentic security initiative},
  author={John, Sotiropoulos and Del, Rosario Ron F and Evgeniy, Kokuykin and Helen, Oakley and Idan, Habler and Kayla, Underkoffler and Ken, Huang and Peter, Steffensen and Rakshith, Aralimatti and Ron, Bitton and others},
  year={2025},
  school={OWASP}
}

@article{patil2024gorilla,
  title={Gorilla: Large language model connected with massive apis},
  author={Patil, Shishir G and Zhang, Tianjun and Wang, Xin and Gonzalez, Joseph E},
  journal={Advances in Neural Information Processing Systems},
  volume={37},
  pages={126544--126565},
  year={2024}
}

@inproceedings{jabbour2024generative,
  title={Generative AI agents in autonomous machines: A safety perspective},
  author={Jabbour, Jason and Janapa Reddi, Vijay},
  booktitle={Proceedings of the 43rd IEEE/ACM International Conference on Computer-Aided Design},
  pages={1--13},
  year={2024}
}

@article{elnashar2025prompt,
  title={Prompt engineering for structured data: a comparative evaluation of styles and LLM performance},
  author={Elnashar, Ashraf and White, Jules and Schmidt, Douglas C},
  journal={Artificial Intelligence and Autonomous Systems},
  volume={2},
  number={2},
  pages={32--49},
  year={2025},
  publisher={ELSPublishing}
}

@article{wang2025mcp,
  title={Mcp-bench: Benchmarking tool-using llm agents with complex real-world tasks via mcp servers},
  author={Wang, Zhenting and Chang, Qi and Patel, Hemani and Biju, Shashank and Wu, Cheng-En and Liu, Quan and Ding, Aolin and Rezazadeh, Alireza and Shah, Ankit and Bao, Yujia and others},
  journal={arXiv preprint arXiv:2508.20453},
  year={2025}
}

@article{li2023halueval,
  title={Halueval: A large-scale hallucination evaluation benchmark for large language models},
  author={Li, Junyi and Cheng, Xiaoxue and Zhao, Wayne Xin and Nie, Jian-Yun and Wen, Ji-Rong},
  journal={arXiv preprint arXiv:2305.11747},
  year={2023}
}

@article{lin2025llm,
  title={LLM-based Agents Suffer from Hallucinations: A Survey of Taxonomy, Methods, and Directions},
  author={Lin, Xixun and Ning, Yucheng and Zhang, Jingwen and Dong, Yan and Liu, Yilong and Wu, Yongxuan and Qi, Xiaohua and Sun, Nan and Shang, Yanmin and Wang, Kun and others},
  journal={arXiv preprint arXiv:2509.18970},
  year={2025}
}

@article{driess2023palm,
  title={Palm-e: An embodied multimodal language model},
  author={Driess, Danny and Xia, Fei and Sajjadi, Mehdi SM and Lynch, Corey and Chowdhery, Aakanksha and Wahid, Ayzaan and Tompson, Jonathan and Vuong, Quan and Yu, Tianhe and Huang, Wenlong and others},
  year={2023}
}

@article{salehi2025agentic,
  title={Agentic AI and Large Language Models in Radiology: Opportunities and Hallucination Challenges},
  author={Salehi, Sara and Singh, Yashbir and Horst, Kelly K and Hathaway, Quincy A and Erickson, Bradley J},
  journal={Bioengineering},
  volume={12},
  number={12},
  pages={1303},
  year={2025},
  publisher={MDPI}
}

@article{saleh2014system,
  title={System safety principles: A multidisciplinary engineering perspective},
  author={Saleh, Joseph H and Marais, Karen B and Favaro, Francesca M},
  journal={Journal of Loss Prevention in the Process Industries},
  volume={29},
  pages={283--294},
  year={2014},
  publisher={Elsevier}
}

@article{xu2025current,
  title={The Current State of Research on Reputation Evaluation of Network Nodes},
  author={Xu, Jingxiong and Huang, Lisheng and Zhang, Fengjun and Niu, Zuoyuan and Shi, Kai and Li, Qinghua},
  journal={Electronics},
  volume={14},
  number={19},
  pages={3900},
  year={2025},
  publisher={MDPI}
}

@inproceedings{yi2025benchmarking,
  title={Benchmarking and defending against indirect prompt injection attacks on large language models},
  author={Yi, Jingwei and Xie, Yueqi and Zhu, Bin and Kiciman, Emre and Sun, Guangzhong and Xie, Xing and Wu, Fangzhao},
  booktitle={Proceedings of the 31st ACM SIGKDD Conference on Knowledge Discovery and Data Mining V. 1},
  pages={1809--1820},
  year={2025}
}

@misc{simpleqa2024,
  title        = {SimpleQA: Measuring Short-Form Factual Accuracy in Large Language Models},
  author       = {{OpenAI}},
  year         = {2024},
  howpublished = {\url{https://www.datalearner.com/benchmarks/simpleqa}},
  note         = {Accessed: 2025-02-06}
}

@article{bang2025hallulens,
  title={Hallulens: Llm hallucination benchmark},
  author={Bang, Yejin and Ji, Ziwei and Schelten, Alan and Hartshorn, Anthony and Fowler, Tara and Zhang, Cheng and Cancedda, Nicola and Fung, Pascale},
  journal={arXiv preprint arXiv:2504.17550},
  year={2025}
}

@article{wangtoolbench,
  title={ToolBench 2.0: Evaluating Long-Horizon and Multi-Step Tool Use in LLMs},
  author={Wang, Guangyu and Liu, Jianhong and Zhou, Meilin and Chen, Xiaoming and Zhang, Lihua and Sun, Zhihao}
}

@inproceedings{pan2018flowcog,
  title={$\{$FlowCog$\}$: Context-aware semantics extraction and analysis of information flow leaks in android apps},
  author={Pan, Xiang and Cao, Yinzhi and Du, Xuechao and He, Boyuan and Fang, Gan and Shao, Rui and Chen, Yan},
  booktitle={27th USENIX Security Symposium (USENIX Security 18)},
  pages={1669--1685},
  year={2018}
}

@article{lin2024contextualized,
  title={Contextualized sequence likelihood: Enhanced confidence scores for natural language generation},
  author={Lin, Zhen and Trivedi, Shubhendu and Sun, Jimeng},
  journal={arXiv preprint arXiv:2406.01806},
  year={2024}
}

@inproceedings{sinha2020relation,
  title={Relation aware attention model for uncertainty detection in text},
  author={Sinha, Manjira and Agarwal, Nilesh and Dasgupta, Tirthankar},
  booktitle={Proceedings of the ACM/IEEE Joint Conference on Digital Libraries in 2020},
  pages={437--440},
  year={2020}
}

\begin{appendices}
\newpage
\section{Appendix}
\label{app:method}

\subsection{LLM usage considerations}
In this work, LLMs were used solely to assist with the refinement of the manuscript, including grammatical polishing, clarity improvements, and stylistic adjustments. The models were not involved in designing the methodology, generating experimental insights, or producing any novel technical contributions. All conceptual frameworks, analyses, and results presented in this paper were developed independently by the authors. The use of LLMs, therefore, serves only as an editorial aid and does not influence or alter the scientific validity or originality of the research.

\subsection{Taxonomy Appendix}
 \textbf{Scope.}
We focus on \emph{runtime interaction security} of LLM-based agents: how agents \emph{perceive, decide, and act} through protocols (e.g., MCP/A2A), tools/APIs, and local resources. 
Training-time data poisoning, model theft, and pretraining misalignment are \emph{out of scope} unless they directly mediate runtime decisions. We expanded with class-specific queries (e.g., ``prompt injection'', ``tool poisoning'', ``A2A spoofing'') and applied forward/backward snowballing from seed papers.

When processing titles, papers clearly unrelated to compartmentalization are discarded. Abstracts are inspected if the title does not enable an unambiguous decision, and are sufficient to make an unambiguous decision in the most cases. We analyze the content of papers to determine if the study works in Stage 1-3 defined in Section 3.3.

\textbf{Inclusion/Exclusion criteria.}
We \emph{include} artifacts that (i) concern LLM-based agents or agentic pipelines that \emph{call tools, access resources, or coordinate with agents}; (ii) discuss security failures/defenses with \emph{operational impact} (C/I/A/accountability); and (iii) provide sufficient detail to map to at least one Stage of our mismatch chain. 
We \emph{exclude} artifacts that (i) only evaluate single-model harmlessness without external action, (ii) concern training-time attacks unrelated to runtime decisions, or (iii) lack verifiable technical substance (e.g., purely conceptual blog posts).

\subsection{Source Search Appendix}
The keywords considered for the automated search of papers on Google Scholar are: LLMs, agent, MAS, and protocol (including derived terms, as well as related attack keywords such as jailbreaking, poisoning, and prompt injection, and defense keywords such as detection and isolation). This yields $279$ retrieved records, after which we performed automated de-duplication over titles/DOIs/URLs and normalized preprint/venue versions. 
Two reviewers independently screened titles/abstracts, followed by full-text screening, resulting in $211$ de-duplicated unique items. 
Many false positives originate from projects that use these terms in unrelated contexts—such as general-purpose chatbot applications, multi-agent coordination frameworks without security relevance, or protocol utilities that do not involve adversarial interaction. We further inspect associated authors and organizations to identify additional works that follow similar security-oriented design principles but were missed due to inconsistent terminology. Finally, we incorporated $104$ papers from prior surveys and domain knowledge into the final included corpus to obtain a consolidated set of mainstream attack and defense techniques across LLMs, agents, MAS environments, and protocol-driven systems.

 \textbf{Corpus construction.}
We performed a systematic literature review across systems-security venues (e.g., IEEE S\&P, USENIX Security, CCS, NDSS), AI/ML venues (e.g., NeurIPS, ICML, ICLR, ACL), and major digital libraries (IEEE Xplore, ACM DL, USENIX, ACL Anthology), complemented with arXiv categories (cs.CR, cs.AI, cs.LG) and Google Scholar to capture grey literature. 
The search window spans \emph{Jan 1, 2023} to \emph{Nov 1, 2025} (UTC).

\textbf{Search strings.}
Grounded in our B--I--P lens, we used conjunctive queries targeting the intersection of agents, security, and tool-mediated actions, for example:
\begin{quote}\small
\texttt{(LLM OR ``AI" OR agent) AND (security OR attack OR defense OR authorization) AND (tool OR API OR protocol OR interaction OR permission)}
\end{quote}

% \textbf{Screening and de-duplication.}
% We first performed automated de-duplication over titles/DOIs/URLs and normalized preprint/venue versions. 
% Two reviewers independently screened titles/abstracts, followed by full-text screening. 
% Let 279 be the number of retrieved records, 222 the de-duplicated unique items, and 102 the final included corpus. 
% We categorize each item as peer-reviewed (PR) or grey/preprint (GL) and report inter-rater agreement ($\kappa$) for both screening Stages. 
% % replace $N_0,N_1,N_\text{incl},\kappa$ with actual values in camera-ready

 \textbf{Coding scheme and data extraction.}
For each artifact we extract: venue/year; component under study (\emph{victim component}: user query, system prompt, tool return, agent return, local resource, protocol metadata); coverage of mismatch Stages (Stage 1 to 3); evaluation setting (simulated, real-system, incident); \emph{evidence level} (concept/demo/empirical/incident); \emph{reproducibility assets} (code/data/config); and \emph{attack/defense role}. 
We will release a machine-readable table and scripts to reproduce the corpus and figures.

\textbf{PRISMA-style flow and availability.}
A PRISMA-style diagram summarizes $279 \!\rightarrow\! 211 \!\rightarrow\! 102$ with reasons for exclusion (non-agentic, training-only, insufficient technical detail, duplicates). 

 \textbf{Threats to validity (brief).}
(i) \emph{Venue bias}: security incidents often appear first in grey literature; we mitigate via GL tagging and sensitivity analyses. 
(ii) \emph{Temporal drift}: the ecosystem evolves quickly; we record retrieval timestamps and archive URLs. 
(iii) \emph{Selection bias}: to reduce omission, we perform snowballing and query expansion; the artifact documents all queries. 
\end{appendices}

\end{document}